\documentclass[sigconf]{acmart}
\usepackage[T1]{fontenc}
\usepackage{graphicx}
\usepackage{xspace}
\usepackage{xcolor}
\usepackage{amsmath}
\usepackage{enumitem}
\usepackage{stfloats}
\usepackage{balance}

\newcommand*{\halo}{\texttt{halo2}\xspace}
\newcommand*{\sn}{\texttt{ZKTorch}\xspace}

\newcommand{\minihead}[1]{{\vspace{.45em}\noindent\textbf{#1.} }}

\AtBeginDocument{%
  }

\begin{document}

\title{\sn: Compiling ML Inference to Zero-Knowledge Proofs via Parallel Proof Accumulation}

\author{Bing-Jyue Chen}
\affiliation{%
 \institution{UIUC}
 \country{USA}}
\email{bjchen4@illinois.edu}

\author{Lilia Tang}
\affiliation{%
  \institution{UIUC}
  \country{USA}}
\email{liliat2@illinois.edu}

\author{Daniel Kang}
\affiliation{%
  \institution{UIUC}
  \country{USA}}
\email{ddkang@illinois.edu}

\begin{abstract}
%The abstract should briefly summarize the contents of the paper in
As AI models become ubiquitous in our daily lives, there has been an increasing demand for transparency in ML services.
However, the model owner does not want to reveal the weights, as they are considered trade secrets.
To solve this problem, researchers have turned to zero-knowledge proofs of ML model inference.
These proofs convince the user that the ML model output is correct, without revealing the weights of the model to the user.

\quad Past work on these provers can be placed into two categories.
The first method compiles the ML model into a low-level circuit, and proves the circuit using a ZK-SNARK.
The second method uses custom cryptographic protocols designed only for a specific class of models.
Unfortunately, the first method is highly inefficient, making it impractical for the large models used today, and the second method does not generalize well, making it difficult to update in the rapidly changing field of machine learning.

\quad To solve this, we propose \sn, an open source end-to-end proving system that compiles ML models into base cryptographic operations called basic blocks, each proved using specialized protocols. \sn is built on top of a novel parallel extension to the Mira accumulation scheme, enabling succinct proofs with minimal accumulation overhead. These contributions allow \sn to achieve at least a $3\times$ reduction in the proof size compared to specialized protocols and up to a $6\times$ speedup in proving time over a general-purpose ZKML framework.

% These basic blocks are easily swapped out and updated when a new machine learning model is designed.
% By directly implementing these basic blocks in zero knowledge, as opposed to compiling down to a low level circuit, we can achieve \todo{}{X speedup} over previous general purpose zkml provers while maintaining concise proof sizes.

\end{abstract}

\maketitle

\section{Introduction}
\label{sec:introduction}
The widespread adoption of AI models, ranging from LLMs for text generation \cite{touvron2023llama} to ResNet-50 \cite{he2016identity} for image classification, has driven demand for transparent and accountable AI \cite{karissa2023what}. However, there is a fundamental challenge of balancing transparency with protecting proprietary model weights. To address this, researchers have turned to zero-knowledge proofs (typically ZK-SNARKs) for ML model inference to perform trustless and weight privacy-preserving audits \cite{waiwitlikhit2024trustless,kang2022scaling,chen2024zkml,lee2020vcnn,weng2022pvcnn,feng2021zen,sun2024zkllm,liu2021zkcnn}. These proofs allow model providers to demonstrate that a model passed an audit without revealing its weights.

ZK-SNARK-based approaches transform ML model inference into cryptographic operations in two general ways. One class of techniques encodes the model's computation as a general arithmetic or logical circuit using a general-purpose ZK-SNARK system \cite{kang2022scaling}, such as \halo \cite{halo2}. Another class designs bespoke cryptographic protocols for specific model types, such as convolutional neural networks (CNNs) or large language models (LLMs) \cite{lee2020vcnn,weng2022pvcnn,sun2024zkllm,liu2021zkcnn}.

Unfortunately, these methods either incur significant prover overhead or lack generality. General-purpose ZK-SNARK systems are not optimized for ML, leading to high overheads. Proving a single inference for a large model like Llama2-7B could take hundreds of hours, making this approach impractical. Bespoke cryptographic protocols, on the other hand, do not support a wide range of modern ML models. For instance, zkCNN does not support LLMs \cite{liu2021zkcnn}, and zkLLM does not support CNNs \cite{sun2024zkllm}.

To resolve this tension, we propose \sn, the first end-to-end proof generation for
every model in the MLPerf Edge Inference suite v4.1 \cite{mlperf}, surpassing the
capabilities of prior ZKML systems. \sn is an end-to-end
system for compiling and proving ML model inference with both
broad coverage and efficiency in mind. To support a wide variety
of layers, \sn adopts a modular design that compiles diverse
ML models into a set of base cryptographic protocols designed
to support common ML operations (referred to as basic blocks).
However, this technique alone produces many small proofs for
each basic block, resulting in large proof sizes.

To address this issue, \sn is built on top of the Mira accumulation scheme \cite{beal2024mira}, a pairing-based recursive SNARK designed for efficient composition of proofs. Mira offers two key advantages over prior work such as ProtoStar \cite{bunz2023protostar} critical to \sn: 1) it supports small recursive proofs for pairing-based arguments which enables scalable verification and succinct proofs and 2) it integrates naturally with efficient cryptographic primitives such as the KZG commitment scheme \cite{kate2010constant} and state-of-the-art table lookup schemes \cite{eagen2022cq} which are often used to prove nonlinearities in prior work \cite{chen2024zkml,sun2024zkllm}. These properties make Mira an ideal foundation for scalable ZKML systems, maintaining succinct proofs with minimal accumulation prover overhead.

% While Mira demonstrates the potential for small proof sizes and low prover overhead, two challenges remain: limited layer coverage and inefficient aggregation of multiple proof instances. In particular, it is not immediately clear how to support a variety of ML layers using only the restricted set of pairing-based protocols available. Moreover, accumulating multiple proof instances sequentially is not efficient enough for large-scale models.

% To address these limitations, \sn adopts a modular design that compiles diverse ML models into a set of base cryptographic protocols (referred to as \emph{basic blocks}), which are low-level primitives implemented using efficient pairing-based techniques. 
However, Mira only provides the accumulation protocol for three basic blocks in \sn. We extend Mira to support pairing-based arguments for more basic blocks to further reap its benefits. Building on this foundation, \sn provides a transpiler that maps high-level ML layers to basic blocks and a compiler that reorganizes models to optimize proving performance. As a result, \sn supports 61 commonly used ML layers, including those found in CNNs, recurrent neural networks (RNNs), and transformers.

To further scale proof generation, we generalize Mira’s accumulation protocol to support parallel aggregation. Mira performs accumulation by calculating a random linear combination between an accumulation instance and a SNARK proof. \sn's accumulator changes a SNARK proof into an accumulation instance, allowing accumulation between two arbitrary accumulation instances. This approach alleviates the bottleneck of sequential recursion, significantly reducing overall proving time by up to $6\times$. Moreover, the proof size of \sn is at least $3\times$ smaller compared to specialized protocols such as Mystique \cite{weng2021mystique} and zkCNN \cite{liu2021zkcnn}.

% Finally, \sn demonstrates how to implement ML operations with substantial gains by leveraging pairing-based cryptographic primitives (e.g., the KZG commitment scheme). For example, vector addition—a key operation in residual connections—can be performed up to $620\times$ faster with direct KZG-based implementations compared to general-purpose ZK-SNARK circuits \cite{halo2}. We also show how to build specialized ML operations (such as \texttt{Max}) by integrating low-level cryptographic primitives.

In summary, \sn makes three key contributions:
1)~it extends the set of base cryptographic protocols to support 61 commonly used ML layers across CNNs, RNNs, and LLMs;
2)~it generalizes Mira’s accumulation protocol for parallel proof aggregation, achieving up to $6\times$ faster proving times and at least $3\times$ smaller proofs compared to specialized schemes; and
3)~it incorporates a carefully designed compiler and transpiler that maximize the use of efficient basic blocks, resulting in up to a $4.8\times$ reduction in proving time and a $2,398\times$ reduction in proof size.

Together, these contributions enable \sn to combine flexibility and efficiency in a single framework, providing scalable zero-knowledge inference for real-world ML models. We open source \sn and provide the link in Appendix A.

\section{Preliminaries}

In this section, we provide the relevant background on important cryptographic primitives, ML models, and how they relate to ZK-SNARKs. We further discuss how specific considerations in ML models
relate to the security of ZK-SNARKs for ML.

We assume $\mathbb{F}$ is a finite prime field, and denote $\mathbb{F}_{<d}[X]$ to be the set of univariate polynomials over $\mathbb{F}$ with degree smaller than $d$. Let \(\mathbb{G}_1\) and \(\mathbb{G}_2\) be two elliptic curves with a pairing 
\(e : \mathbb{G}_1 \times \mathbb{G}_2 \rightarrow \mathbb{G}_T\). 
Let \(p\) be the order of \(\mathbb{G}_1\) and \(\mathbb{G}_2\), and \(G\) and \(H\) be generators of 
\(\mathbb{G}_1\) and \(\mathbb{G}_2\). We use the notation
$[x]_1 = xG \in \mathbb{G}_1, [x]_2 = xH \in \mathbb{G}_2$

\subsection{Polynomial Commitment}
A polynomial commitment scheme \cite{kate2010constant} allows a prover to commit to a polynomial $f \in \mathbb{F}_{\le d}[X]$.
% A polynomial commitment scheme also allows the prover to prove that the committed polynomial evaluates to $f(x)$ at any point $x \in \mathbb{F}$.
A polynomial commitment scheme $\mathcal{C}$ consists of the following algorithms:
\begin{itemize}[leftmargin=*]
\item $\textsf{pp} \leftarrow \textsf{setup}(1^\lambda, d)$, where \textsf{pp} are public parameters to commit to a polynomial of degree at most $d$.
\item $\textsf{com} \leftarrow \textsf{commit}(\textsf{pp}, f, r)$, where \textsf{com} is a commitment to a polynomial $f \in \mathbb{F}_{\le d}[X]$ with randomness $r$.
\item $(\pi, y) \leftarrow \textsf{open}(\textsf{pp}, f, x, r)$, where $\pi$ is a proof proving $f(x) = y$.
\item $\{0, 1\} \leftarrow \textsf{verify}(\textsf{pp}, \textsf{com}, x, y, \pi)$.
\end{itemize}

In this work, we use the KZG polynomial commitment scheme \cite{kate2010constant}. 
The KZG commitment scheme makes the SDH assumption \cite{boneh2004short}. The SDH assumption states that in a group $\mathbb{G}$ of prime order $p$ states that given $g, g^x, g^{x^2}, \ldots, g^{x^q} \in \mathbb{G}$, we cannot output $(c, g^{1/(x+c)})$ where $c \in \mathbb{Z}_p$ in polynomial time.

The KZG commitment scheme relies on a structured reference string $\texttt{srs}=[\tau^0]_1,\ldots,[\tau^n]_1$ for some uniform trapdoor $\tau\in\mathbb F$.
This structured reference string is generated by group of parties via a powers of tau ceremony as described in \cite{cryptoeprint:2017/1050}.
Once the \texttt{srs} is generated, as long as one of the parties is honest, no party can recover the value of the trapdoor $\tau$ \cite{cryptoeprint:2017/1050}. The security properties of KZG commitment scheme when using in ZK-SNARKs are often proved with the algebraic group model (AGM), which requires any adversary that outputs a group element $c$ to return the corresponding sequence of field elements $s_i$ such that $c=
    \sum_i g^{x^i\cdot s_i}$ \cite{gabizon2019plonk}.

KZG computes the commitment of a polynomial $f$ by $\textsf{com}=g^{f(\tau)}=[f(\tau)]_1$. When opening $f$ at $x\in\mathbb{F}$, the prover returns $y=f(x)\in\mathbb{F}$ and the corresponding proof $\pi=[q(\tau)]_1$, where $q(X)=\frac{f(X)-y}{X-x}$. The verifier then checks if the pairings $e(\pi, [\tau-x]_2)$ and $e(\textsf{com}-[y]_1, [1]_2)$ are equal. Moreover, we can batch the opening proofs for $f_0(X), f_1(X), ..., f_n(X)$ if we evaluate them at the same point $X=x$. To enable this, the verifier sends a random $\gamma\in\mathbb{F}$ right after receiving the openings $f_0(x), f_1(x), ..., f_n(x)$ from the prover. Then, the prover computes the batch proof $\pi=[h(\tau)]_1$, where $h(X)=\frac{\sum_{i=0}^n\gamma^i (f_i(X)-f_i(x))}{X-x}$. Finally, the verifier checks if $e(\pi, [\tau-x]_2) 
    = \,e\left(\sum_{i=0}^n \gamma^k (\mathsf{com}_i - [f_i(x)]_1), [1]_2\right)$.

\subsection{Interactive Proofs and ZK-SNARKs}
An interactive proof $\Pi = (\mathcal{P}, \mathcal{V})$ for a relation $\mathcal{R}$ consists of an interactive protocol where the prover $\mathcal{P}$, who holds a witness $w$, interacts with the verifier $\mathcal{V}$ on public input $\mathsf{pi}$ to generate a proof string $\pi$ to convince $\mathcal{V}$ that $(\mathsf{pi}, w) \in \mathcal{R}$.

Building on top of interactive protocols, ZK-SNARKs are a class of cryptographic protocols that provide the following
properties: 1) zero-knowledge, 2) non-interactivity, 3) succinctness, 4)
knowledge soundness, and 5) completeness \cite{petkus2019and}. Informally, the succinctness refers to the fact that the proof produced are small in size and can be verified quickly and the non-interactivity is achieved through Fiat-Shamir transform \cite{Attema2021fiatshamir} over a compatible interactive protocol. 

The security properties in this work are defined as follows:

\noindent\textbf{Perfect completeness.} For every $(\mathsf{pi}, w) \in \mathcal{R}$,
  \[
  \Pr[
  \mathcal{V}(\mathsf{pi}, \mathcal{P}(\mathsf{pi}, w)) = \text{accept}] = 1
  \]

\noindent\textbf{Knowledge soundness.} For any probabilistic polynomial time (PPT) prover $\mathcal{P}^*$, there exists a PPT extractor $\mathcal{E}$ such that given access to the entire executing process and the randomness of $\mathcal{P}^*$, $\mathcal{E}$ can extract a witness $w$ such that $\pi^* \leftarrow \mathcal{P}^*(\mathsf{pi}, w)$, and $w \leftarrow \mathcal{E}^{\mathcal{P}^*}(\mathsf{pi}, \pi^*)$, the following inequality holds
  \[
  \Pr[(\mathsf{pi}, w) \not\in \mathcal{R} \land \mathcal{V}(\mathsf{pi},\pi^*)=\text{accept}] \leq \text{negl}(\lambda)
  \]

\noindent\textbf{Zero knowledge.} Let $\langle \mathcal{P}, \mathcal{V}^* \rangle$ denote the set of all messages in the protocol (i.e., transcript). There exists a PPT simulator $\mathcal{S}$ such that for any PPT algorithm $\mathcal{V}^*$, $(\mathsf{pi}, w) \in \mathcal{R}$,
  $\text{View}\left(\langle \mathcal{P}, \mathcal{V}^* \rangle\right) \approx \mathcal{S}^{\mathcal{V}^*}(\mathsf{pi})$ holds.
Note that the simulator $\mathcal S$ has access to the trapdoor $\tau$ of the KZG commitment scheme, which allows the simulator to generate valid proofs without access to the witness $w$.
\vspace{5pt}

There are many
ways to construct ZK-SNARKs, ranging from general-purpose
circuits \cite{halo2} to specialized protocols for specific operations, such as
ML models \cite{liu2021zkcnn} or identity protocols \cite{pauwels2022zkkyc}.

Although protocols vary in their implementations, all ZK-SNARK protocols operate
over finite fields and convert the computation to prove into an intermediate
representation that is compiled into cryptographic operations. For example, the
\halo protocol uses a 256-bit field (several are admissible) and a randomized
arithmetic intermediate representation (AIR) with preprocessing to represent
computation \cite{gabizon2021from}. Other proving systems use 64- or 32-bit
fields, and other intermediate representations such as arithmetic circuits.

Thus, there are two critical properties for ZK-SNARKs: the security of
the underlying cryptographic protocol and the correctness of the intermediate representation. 

\subsection{Accumulation Schemes} 
\label{prelim:acc}
An accumulation scheme is an approach to fold multiple proofs into one, which reduces the proof size and the verifier's overhead. 

We use the definition for an accumulation scheme from \cite{bunz2023protostar}, repeated here. An accumulation scheme for a NARK (e.g., a basic block in \sn) $(\mathcal{P}_{\mathsf{NARK}}, \mathcal{V}_{\mathsf{NARK}})$ is a triple of algorithms 
$\text{acc} = (\mathcal{P}_{\text{acc}}, \mathcal{V}_{\text{acc}}, \mathcal{D})$, all of which have access to the same random 
oracle $\rho_{\text{acc}}$ as well as $\rho_{\mathsf{NARK}}$, the oracle for the NARK. The algorithms 
have the following syntax:

\begin{itemize}[leftmargin=*]
    \item $\mathcal{P}_{\text{acc}}(\mathsf{pi}, \pi = (\pi.x, \pi.w), \text{acc} = (\text{acc}.x, \text{acc}.w)) 
    \\ \rightarrow \{ \text{acc}' = (\text{acc}'.x, \text{acc}'.w), \mathsf{pf} \}$. The accumulation 
    prover $P_{\text{acc}}$ takes as input a statement $\mathsf{pi}$, NARK proof $\pi$, and accumulator 
    $\text{acc}$ and outputs new accumulator $\text{acc}'$ and correction terms $\mathsf{pf}$.

    \item $\mathcal{V}_{\text{acc}}(\mathsf{pi}, \pi.x, \text{acc}.x, \text{acc}'.x, \mathsf{pf}) \rightarrow v$. The 
    accumulation verifier takes as input the statement $\mathsf{pi}$, the instances of the NARK proof, the old 
    and new accumulator, the correction terms, and accepts by outputting 0 and rejects otherwise.

    \item $\mathcal{D}(\text{acc}) \rightarrow v$. The decider on input $\text{acc}$ accepts by outputting 0 
    and rejects otherwise.
\end{itemize}
Note that $x$ and $w$ in the above definition can be understood as the public and private parts of the instances.

Mira \cite{beal2024mira} enables a pairing-based SNARK prover to perform accumulation  on elliptic group elements (i.e., the proofs and the commitments of witnesses) without needing to perform accumulation on both plaintext field elements and their commitments as required in previous protocols such as ProtoStar \cite{bunz2023protostar}. Specifically, Mira performs accumulation by calculating a random linear combination between an accumulation instance and a SNARK proof (and the corresponding input/output commitments used in the pairing check of the SNARK). Mira also shows how to derive the error when performing the elliptic pairing check on the accumulation instance.

We apply Mira to fold multiple proofs from CQ and CQLin into one. Since there are other basic blocks in \sn also based on low-degree algebraic tests, we extend the accumulation scheme for them. We provide a high-level introduction to the Mira accumulation scheme \cite{beal2024mira} here. We use $x=\mathsf{Enc}(\mathsf{pk}, \hat{x})$ to represent KZG commitment, where $x\in\mathbb{G}$ and $\hat{x}\in\mathbb{F}^d$ are the commitment and the original data.  Let
$\mathsf{V}_{\text{NARK}}$ be the NARK verifier. The NARK proof has instance $\pi.\mathsf{x} = [C_i]_{i=1}^k$
and witness $\pi.\mathsf{w} = \{[m_i]_{i=1}^k\}$. The accumulator instance is denoted by $\mathsf{acc}.\mathsf{x} = \{\mathsf{pi}, [C_i]_{i=1}^k, [r_i]_{i=1}^{k-1}, E, \mu\}$. The witness is denoted by $\mathsf{acc}.\mathsf{w} = \{[m_i]_{i=1}^k, e\}$.

Following \cite{beal2024mira}, given that the degree-$j$ terms in the algebraic test of the original NARK is $f_{j}^{\mathsf{V_{\text{lin}}}'}([\hat{m}_i]_{i=1}^k, [r_i]_{i=1}^{k-1})$. 
  The relaxed algebraic test is in the following form:
  \[f^{\mathsf{V_{\text{lin}}}}([\hat{m}_i]_{i=1}^k, [r_i]_{i=1}^{k-1}) =-\hat{e} + \sum_{j=0}^d\mu^{d-j} \cdot f_{j}^{\mathsf{V_{\text{lin}}}'}([\hat{m}_i]_{i=1}^k, [r_i]_{i=1}^{k-1})=0\] Mira defines the accumulated predicate as follows:

\begin{itemize}[leftmargin=*]
  \item Challenge derivation is valid: \\ $r_i = \rho_{\text{NARK}}(r_{i-1}, C_i) \, \forall i \in [k-1], r_0 = \rho_{\text{NARK}}(\mathsf{pi})$
  \item Commitment openings are valid: $\mathsf{Commit}(\mathsf{ck}, m_i) = C_i \, \forall i \in [k]$
  \item Given $\mathsf{V}_{\text{lin}}$ is the pairing checks to verify $f^{\mathsf{V_{\text{lin}}}}\overset{?}{=}0$, the relaxed algebraic test holds ($\ell$ denotes the number of the algebraic tests needed in NARK): $\mathsf{V}_{\text{lin}}(\mathsf{pi}, [m_i]_{i=1}^k, [r_i]_{i=1}^{k-1}, \mu, e) = 0^\ell$
\end{itemize}

$\mathcal{P}_{\text{acc}}$ first derives the random challenges in the NARK and computes new error terms based on Sec 3.6 in \cite{bunz2023protostar}. Then $\mathcal{P}_{\text{acc}}$ samples randomness $\gamma$ to update the new error $e$ and the corresponding $E$. Next, $\mathcal{P}_{\text{acc}}$ performs a random linear combination (RLC) for the tuple $(\mathsf{pi},[C_i]^k_{i=1},[r_i]^{k-1}_{i=1},[m_i]^k_{i=1})$ with $\pi$ and $\text{acc}$ to update $\text{acc}$. $\mathcal{V}_{\text{acc}}$ checks if the following are true: 1) the randomness derivation is valid, 2) the error terms are updated correctly, and 3) the RLC is correct. $\mathcal{D}$ accepts the $\text{acc}$ if the relaxed algebraic test $\mathsf{V}_{\text{lin}}$ holds. The checks for commitment openings can be omitted because Mira uses trivial identity commitment (i.e., $\mathsf{Commit}$ returns the input itself, which means $[C_i]=[m_i],\; \forall i$ and $E=e$).

Relying on the homomorphism of the underlying commitment scheme, the accumulation prover can efficiently fold proofs. However, this formulation constrains us to perform accumulation sequentially. Given that there are $N$ proofs to fold, the time complexity will be $O(N)$. To improve this time complexity, we view $\pi$ as a special case of $\text{acc}$ and generalize the above definition in Sec 4.4 to enable parallel accumulation.

\subsection{ML Models and ZK-SNARKs}

ML models consist of highly structured computation that are typically described
as \emph{layers} that take as input one or more \emph{tensors}. These layers
include operations ranging from matrix multiplication to the ReLU non-linearity
\cite{lecun2015deep}. Some of the tensors are fixed ahead of time (i.e., are
\emph{weights}) and others must be computed dynamically depending on the input.
\cite{lecun2015deep} provides a detailed description of ML model computation
for further reading.

ML model computation is natively computed in floating-point arithmetic, which is
not natively representable in finite fields. To address this issue, ZK-SNARK
systems for ML approximate the floating-point values with fixed-point
arithmetic, where a number is represented by a value and a scale factor
\cite{chen2024zkml}.

ML models vary in their computational patterns, but within a model,
computational patterns are often repeated. For example, the widely used ResNet
series alternates between convolutions, residual layers, and ReLU
\cite{he2016identity}. LLMs use feed-forward networks (matrix
multiplications) and attention heads (matrix multiplication, point-wise
multiplication, non-linearities, and the softmax) \cite{touvron2023llama}.

A key insight we leverage to construct \sn is that the number of
unique operations used by common ML models is relatively small. Consider the widely
used MLPerf inference benchmark \cite{reddi2020mlperf}, which covers a wide
range of ML models. Across all of the models in the edge category, only 61
operations are used. These operations further cover other popular models, such
as the Llama LLM series.

\minihead{Lookup Arguments}
Lookup arguments are NARKs for proving set inclusion. Given a table $T$ known to both the prover and verifier, the prover can prove that all elements of a committed vector are in $T$. Table sizes to support scaling and non-linearities are often much larger than the committed vector size, so we use the cq protocol \cite{eagen2022cq} because it is compatible with the KZG commitment scheme and has proving time independent of the table size.

\section{Overview of \sn}
\label{sec:overview}

\begin{figure*}[t]
    \centering
    \includegraphics[width=.96\textwidth]{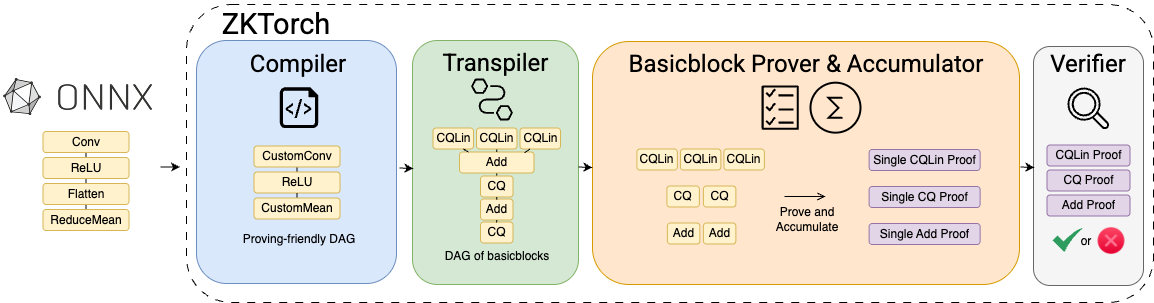}
    \caption{Architecture diagram of \sn}
    \label{fig:arch}
\end{figure*}

\sn is designed to bridge the gap between broad coverage of ML layers and efficient proof generation, based on the key insights and contributions described in Section~\ref{sec:introduction}. In particular, \sn extends pairing-based cryptographic techniques to support 61 common layers across CNNs, RNNs, and transformers, and  generalizes Mira’s accumulation protocol to enable parallel proof aggregation with minimal overhead. With a carefully optimized compiler and transpiler, \sn provides an end-to-end framework for scalable zero-knowledge proofs of real-world ML model inference.

At a high level, \sn compiles an ONNX model graph, consisting of various ML layers and operations, into a directed acyclic graph (DAG) of \emph{basic blocks}, which are low-level zero-knowledge primitives tailored to efficiently handle common ML operations while preserving weight or input privacy. We illustrate this architecture in Figure \ref{fig:arch}. \sn leverages
\begin{itemize}[leftmargin=*] \item \textbf{an optimizing compiler}  to reorder and combine model layers to prove them with less overhead,
% It fuses compatible operations to reduce both proving time and proof size.
\item \textbf{a transpiler} to automatically translate ML layers in the ONNX graph into compositions of these basic blocks, and
\item \textbf{a set of cryptographic primitives} based on KZG polynomial commitments \cite{kate2010constant} implemented using arkworks libraries \cite{arkworks} to support a diverse set of ML layers and parallel proof aggregation by extending Mira \cite{beal2024mira}.
% each implemented using pairing-based techniques (e.g., KZG \cite{kate2010constant} commitments) and extended from Mira \cite{beal2024mira} to support more ML layers and parallel proof aggregation.
\end{itemize}

\minihead{Witness generation and encoding} To prove correctness of inference without revealing internal weights or inputs, \sn adopts a witness-generation process reminiscent of traditional model inference. This process scales the relevant tensors to elements of a finite prime field and encodes them by interpreting their final dimension as polynomials. The prover commits to them using KZG for each tensor (input, output, and intermediate activations).

\minihead{Layer-by-layer proving} With these commitments, \sn proves layer-by-layer that each basic block within the layer was computed correctly.
Each basic block's output commitments become inputs to subsequent blocks in the DAG, and \sn maintains intermediate proofs to ensure the correctness of every step.

\minihead{Parallel proof accumulation} Naively proving and verifying every basic block in sequence would lead to large proof sizes and high overall overhead. To address this, \sn supports \emph{parallel} accumulation. After generating proofs for the basic blocks, \sn aggregates them—often batching proofs for the same block type—into a single succinct proof. This generalization of Mira’s accumulation protocol significantly reduces total proving time (by up to $6\times$ in our measurements) while maintaining lightweight proof sizes (at least $3\times$ smaller than specialized protocols).

\minihead{Scalability and full coverage} Finally, \sn demonstrates end-to-end proofs of entire models on a single server, rather than focusing on just one or two specialized network topologies. Thanks to its comprehensive set of basic blocks and optimized compilation pipeline, \sn can handle all models in the MLPerf Inference: Edge v4.1 suite~\cite{mlperf}, including CNNs, RNNs, and transformer-based architectures. In doing so, \sn illustrates for the first time how a general-purpose ZKML system can combine flexibility and efficiency for modern, real-world ML models.
\section{Basic Blocks}
\begin{table}[t!]

\centering
\caption{List of basic blocks rules implemented in \sn. \checkmark indicates that the accumulation scheme is supported.}
\label{table:basicblocks}
\begin{tabular}{l|l|c}
Basic Block & Category & Accumulation \\
\hline
Add & Arithmetic Operations & \checkmark \\
Div & Arithmetic Operations & \\
Mod & Arithmetic Operations & \\
Mul & Arithmetic Operations & \checkmark \\
MulConst & Arithmetic Operations & \checkmark \\
MulScalar & Arithmetic Operations & \checkmark \\
Sub & Arithmetic Operations & \checkmark \\
Sum & Arithmetic Operations & \checkmark \\
BooleanCheck & Boolean Operations & \\
Eq & Boolean Operations & \checkmark \\
MaxProof & Max & \\
Concat & Shape Operations & \checkmark \\
CopyConstraint & Shape Operations & \\
Permute & Shape Operations & \checkmark \\
CQ & Tables & \checkmark \\
CQ2 & Tables & \checkmark \\
CQLin & Matrix Multiplication & \checkmark \\
MatMul & Matrix Multiplication & \checkmark \\
OneToOne & Sorting Checks & \\
Ordered & Sorting Checks & \\
\end{tabular}

\end{table}

The first major component of \sn are basic blocks. They are selected to prove common operations in ML model layers.

We show a list of basic blocks in Table~\ref{table:basicblocks} along with a categorization
of the kinds of basic blocks. These include arithmetic operations (e.g., vector
addition), matrix multiplication, tables (e.g., pointwise non-linearities), and
shape operations (e.g., permutations). Among all the basic blocks, we can support parallel accumulation for 13 basic blocks. In our compiler and transpiler, we optimize our layer implementations to avoid using the remaining seven basic blocks, which only appear in 18 infrequent layers out of the 61 total layers.

In this section, we first provide the high-level description for each category of basic block. We then argue that the security properties are retained through composition. To effectively fold proofs from several basic blocks into a lightweight final proof, we show the generalized accumulation scheme and its security properties. For each basic block supporting accumulation, we introduce how we relax the algebraic test to make the proof foldable. For other basic blocks, we show the techniques to prove these special operations.
%Of our 20 basic blocks, 4 are novel and 2 are modifications of previous protocols. The remaining 12 are from prior ZK-SNARK protocols. Specifically, the 4 new basic blocks are extensions based on a polynomial interactive oracle proof (IOP), which we will introduce in the following subsection. And the two modification protocols are CQ \cite{eagen2022cq} and CQLin \cite{eagen2023cqlin} because the plain CQ only supports checks for set inclusion and the plain CQLin only supports square matrix multiplications. We modify them to support checks for table lookup and matrix multiplication for arbitrary shapes. %Besides, we also made them zero knowledge. %The modified version of the two protocols is provided in appendix.
\subsection{Basic Block Types}

\minihead{Arithmetic and Boolean Operations}
These are used in arithmetic and Boolean layers and for intermediate computations in computational layers.

\minihead{Max} This basic block is used for Max, Min, and MaxPool layers.

\minihead{Shape Operations}
The concat, copy constraint, and permute basic blocks prove shape operations. We perform shape operations in shape layers such as Concat and Transpose, as well as in computational layers that require inputs to be transformed before applying a basic block. For example, we may need to transpose an input to the MatMul layer before performing a matrix multiplication.

The concat basic block is used for concatenation, while the permute basic block is used for many reshape or transpose operations.
% When these operations do not affect the contents of the last dimension, these operations have no proving cost because they do not require us to change vector commitments. 
In special cases, we use the copy-constraint basic block to prove that every element in the output comes from a value in the input.
% However, it is more expensive than the other basic blocks, so we avoid using it when possible.

\minihead{Tables}
We implement the CQ and CQ2 basic blocks based on the CQ protocol \cite{eagen2022cq}.
CQ checks for set inclusion, while CQ2 checks for lookup table inclusion. The set and lookup table must be known at compile time. A common use case of CQ includes checking that all values of a committed vector are nonnegative. For example, to prove the Div layer, we must check that the resulting remainders after division are nonnegative.
CQ2 is used to prove nonlinearities and scale factor correction after multiplication and division operations.

\minihead{Matrix Multiplication}
We use the CQLin and MatMul basic blocks for proving matrix multiplication.
CQLin implements the CQLin protocol \cite{eagen2023cqlin} which optimizes matrix multiplication when one matrix is fixed, such as in convolutions in CNNs. Since the matrix is known at compile time, it can be processed during setup to enable a proving time linear in the input matrix size. MatMul proves matrix multiplication generally and is based on the Aurora protocol \cite{bensasson18aurora}, and is used for many LLM and RNN matrix multiplications.

\minihead{Sorting Checks}
We support the OneToOne and Ordered basic blocks to check that there is a one-to-one mapping between the input and output and that the input is sorted, respectively. These are currently only used for the ArgMax and TopK layers.

\subsection{Parallel Accumulation}
We provide a formal description of the accumulation scheme, generalizing the method in Sec. 3.3 of \cite{beal2024mira} to support parallel accumulation in \sn. We build upon the Mira accumulation scheme, which only supports sequential accumulation. Specifically, we replace the $\pi$ in base Mira with an $\text{acc}$ instance, and show that $\pi$ is a special case of $\text{acc}$. This method allows \sn to perform accumulation between two arbitrary $\text{acc}$ instances instead of requiring one of them to be $\pi$, allowing for parallel accumulation with techniques such as Merkle tree reduction (i.e. aggregate instances at each level of the tree in parallel).

% \minihead{Protocol} 
We use the notation described in Sec. \ref{prelim:acc} to describe the generalized Mira protocol. In our formulation, the NARK proof $\pi$ is simply a special case of the accumulator, in which $\mu=1$ and $e=0^\ell$. We use an additional bit $b$ to indicate whether an instance is a NARK proof $\pi$ ($b=1$) or not ($b=0$).

\minihead{Accumulation Prover $\mathcal{P}$}
Given commitment key $\mathsf{ck}$ and accumulators $\mathsf{acc}$, $\mathsf{acc'}$,
the accumulation prover $\mathcal{P}_{\text{acc}}^{\rho_{\text{acc}}, \rho_{\text{NARK}}}$ works as follows:

\begin{enumerate}[leftmargin=*]
  \item For each accumulation instance, if its $b$ equals $0$, skip this step. Otherwise, $\mathcal{P}$ derives challenges: $r_i \gets \rho_{\text{NARK}}(r_{i-1}, C_i)$ for all integers $i \in [0,k-1]$, with $r_0 := \rho_{\text{NARK}}(\mathsf{pi})$ and sets the randomness in the instance to be $[r_i]^{k-1}_{i=0}$.
  \item Compute error terms $[e_j]_{j=1}^{d-1}$ such that $e_j = \mathsf{Enc}(\mathsf{pk}, \hat{e}_j) \forall$ integers $j\in[0,d-1]$, $\mathsf{acc.w}.e = \mathsf{Enc}(\mathsf{pk}, \hat{e})$, $\mathsf{acc'.w}.e = \mathsf{Enc}(\mathsf{pk}, \hat{e}')$.
  % \[
  %   \sum_{j=0}^d (X\cdot\mu + \mu')^{d-j} \cdot f_{j}^{\mathsf{V_{\text{lin}}}'}([X \cdot \hat{m}_i + \hat{m}'_i]_{i=1}^k, [X \cdot r_i + r'_i]_{i=1}^{k-1}) = 
  % \]
  \begin{align*}
      & \sum_{j=0}^d (X\cdot\mu + \mu')^{d-j} \cdot f_{j}^{\mathsf{V_{\text{lin}}}'}([X \cdot \hat{m}_i + \hat{m}'_i]_{i=1}^k, [X \cdot r_i + r'_i]_{i=1}^{k-1}) \\
      = &\sum_{j=0}^d \mu'^{d-j} \cdot f_{j}^{\mathsf{V_{\text{lin}}}'}([\hat{m}'_i]_{i=1}^k, [r'_i]_{i=1}^{k-1}) + \sum_{j=1}^{d-1} \hat{e}_j X^j \\
      & \quad + X^d \cdot \sum_{j=0}^d \mu^{d-j}\cdot f_{j}^{\mathsf{V_{\text{lin}}}'}([\hat{m}_i]_{i=1}^k, [r_i]_{i=1}^{k-1}) \\
      = \,&\hat{e}' + \sum_{j=1}^{d-1} \hat{e}_j X^j + \hat{e}X^d
  \end{align*}

  \item Compute committed error terms: $E_j \gets \mathsf{Commit}(\mathsf{ck}, e_j)$
  \item Derive $\gamma \gets \rho_{\text{acc}}(\mathsf{acc}.\mathsf{x}, \mathsf{acc'}.\mathsf{x}, [E_j]_{j=1}^{d-1}) \in \mathbb{F}$
  \item Compute RLCs:
  \[
    v := (\mu, \mathsf{pi}, [r_i]_{i=1}^{k-1}, [C_i]_{i=1}^k, [m_i]_{i=1}^k)
  \]
  \[
    v' := (\mu', \mathsf{pi}', [r_i']_{i=1}^{k-1}, [C_i']_{i=1}^k, [m_i']_{i=1}^k)
  \]
  \[
    v'' := (\mu'', \mathsf{pi}'', [r_i'']_{i=1}^{k-1}, [C_i'']_{i=1}^k, [m_i'']_{i=1}^k) \gets \gamma \cdot v + v'
  \]
  \[
    E'' := E' + \sum_{j=1}^{d-1} \gamma^j \cdot E_j + \gamma^d\cdot E, \,
    e'' := e' + \sum_{j=1}^{d-1} \gamma^j \cdot e_j + \gamma^d\cdot e
  \]
  \item Set new accumulator:
    $\mathsf{acc''}.\mathsf{x} = \{\mathsf{pi}'', [C_i'']_{i=1}^k, [r_i'']_{i=1}^k, E'', \mu''\},\\
    \mathsf{acc''}.\mathsf{w} = \{[m_i'']_{i=1}^k, e''\}$.
  \item Set correction terms: $\mathsf{pf} = [E_j]_{j=1}^{d-1}$
\end{enumerate}

\minihead{Accumulation Verifier $\mathcal{V}$}
\sloppy Given two accumulator instances
$\mathsf{acc}.\mathsf{x} = \{\mathsf{pi}, [C_i]_{i=1}^k, [r_i]_{i=1}^k, E, \mu\}$, $\mathsf{acc'}.\mathsf{x} = \{\mathsf{pi}', [C_i']_{i=1}^k, [r_i']_{i=1}^k, E', \mu'\}$, correction terms $\mathsf{pf}$,
and updated accumulator instance $\mathsf{acc''}.\mathsf{x} = \{\mathsf{pi}'', [C_i'']_{i=1}^k, [r_i'']_{i=1}^k, E'', \mu''\}$:

\begin{enumerate}[leftmargin=*]
  \item For each accumulation instance, if its $b$ equals $0$, skip this step. Otherwise, $\mathcal{V}$ derives challenges: $r_i \gets \rho_{\text{NARK}}(r_{i-1}, C_i)$ for all integers $i \in [0,k-1]$, with $r_0 := \rho_{\text{NARK}}(\mathsf{pi})$ and checks if the randomness in the instance is equivalent to $[r_i]^{k-1}_{i=0}$.
  \item Derive $\gamma \gets \rho_{\text{acc}}(\mathsf{acc}.\mathsf{x}, \mathsf{acc'}.\mathsf{x}, \mathsf{pf})$
  \item Compute $v = (\mu, \mathsf{pi}, [r_i]_{i=1}^{k-1}, [C_i]_{i=1}^k)$,
        $v' = (\mu', \mathsf{pi}', [r_i']_{i=1}^{k-1}, [C_i']_{i=1}^k)$
  \item Let $v'' =\mathsf{acc''}.\mathsf{x}.(\mu'', \mathsf{pi}'', [r_i'']_{i=1}^{k-1}, [C_i'']_{i=1}^k)$. Check $v'' \overset{?}{=} \gamma \cdot v + v'$ and
    $\mathsf{acc''}.\mathsf{x}.E'' \overset{?}{=} \mathsf{acc}'.\mathsf{x}.E' + \sum_{j=1}^{d-1} \gamma^j \cdot E_j + \gamma^d\cdot \mathsf{acc}.\mathsf{x}.E$
\end{enumerate}

\minihead{Accumulation Decider $\mathcal{D}$}
Given instance
$\mathsf{acc}.\mathsf{x} = \{\mathsf{pi}, [C_i]_{i=1}^k, \allowbreak [r_i]_{i=1}^{k-1}, \allowbreak E, \mu\}$ and witness
$\mathsf{acc}.\mathsf{w} = \{[m_i]_{i=1}^k, e\}$,
$\mathcal{D}$ checks that \\ $\mathsf{V}_{\text{lin}}(\mathsf{pi}, [m_i]_{i=1}^k, [r_i]_{i=1}^{k-1}, \mu, e) \overset{?}{=} 0^\ell$.

\minihead{Completeness}
    If the prover $\mathcal{P}$ has two instances of accumulators that can pass the verifier and decider checks, the aggregated instance will also be accepted with probability $1$ following the homomorphism of the commitment scheme.
    
\minihead{Knowledge soundness}
We provide the proof of knowledge soundness in Appendix C. The high-level idea is that we replace the algebraic tests in our generalized Mira with the algebraic checks defined in \cite{bunz2023protostar}, which allows us to use similar techniques to prove that the protocol is knowledge sound. Then we use the techniques in \cite{beal2024mira} to argue that our protocol retains knowledge soundness.

\minihead{Zero knowledge} The simulator $\mathcal{S}$ is constructed as follows:
    \begin{enumerate}[leftmargin=*]
\item $\mathcal{S}$ samples $\{\mathsf{pi}, [C_i]_{i=1}^k\}\leftarrow\mathbb{G}^{k+1}$ and $\mu\leftarrow\mathbb{F}$, uses a random oracle to get $[r_i]_{i=1}^{k-1}$, and calculates $E = \mathsf{V}_{\text{lin}}(\mathsf{pi}, [C_i]_{i=1}^k, \allowbreak [r_i]_{i=1}^{k-1}, \mu, 0^\ell)$. Then it constructs $\mathsf{acc}.\mathsf{x}$, where
\[
    \mathsf{acc}.\mathsf{x} = \{\mathsf{pi}, [C_i]_{i=1}^k, [r_i]_{i=1}^k, E, \mu\}.
\] Similarly, the simulator can sample and calculate
\[
    \mathsf{acc}'.\mathsf{x} = \{\mathsf{pi}', [C_i']_{i=1}^k, [r_i']_{i=1}^k, E', \mu'\} .
\]
Note that the above two equations are true because Mira uses trivial identity commitment as its commitment scheme.

\item $\mathcal{S}$ calculates $[E_j]_{j=1}^{d-1}$ through equations in Sec 3.6 of \cite{bunz2023protostar}.
\item The verifier provides $\gamma$.
\item $\mathcal{S}$ computes the new accumulator using the RLCs introduced in Step 5 of the accumulation prover.
\item $\mathcal{S}$ sends the new accumulator and correction terms $[E_j]_{j=1}^{d-1}$.
\item Given a random $s\in\mathbb{F}$ and a group element $H\in\mathbb{G}$, since we blind our base KZG commitment scheme by adding $sH$ to it. Given that the transcript between the simulator and the adversary and that between the prover and the adversary are indistinguishable, the accumulation scheme satisfies zero knowledge.
\end{enumerate}

\subsection{Security of Composing Basic Blocks}

In the remainder of this section, we first show that composing basic blocks
satisfying certain properties remains secure. We then prove the security of our
individual basic blocks.

We represent the computation of a model inference as a directed acyclic
graph (DAG), where vertices are basic blocks and edges denote where the inputs
flow. We use the \sn compiler and transpiler to transform the original ONNX graph into a \sn DAG that consists of basic blocks.
% Figure \ref{fig:bb1} shows two transformations for GeLU and MatMul.

% Our zero knowledge proofs rely on a structured reference string $\texttt{srs}=[x^0]_1,\ldots,[x^n]_1$ for some uniform trapdoor $x\in\mathbb F$.
% This structured reference string is generated by group of parties via a powers of tau ceremony as described in \cite{cryptoeprint:2017/1050}.
% Once the \texttt{srs} is generated, as long as one of the parties is honest, no party can recover the value of the trapdoor $x$ \cite{cryptoeprint:2017/1050}.

%We require the following properties for each basic block:

%Our goal is to show that the DAG satisfies these three properties. 

All our protocols are complete and knowledge
sound.
% Composition follows directly from the definition. 
If any node in the DAG is invalid but
accepted, the entire DAG is invalid. Let the probability of node $v_i$ being
unsound be $e_{v_i}$. Then, the probability of the DAG being invalid is at most
$\sum e_{v_i}$ by the union bound. 

Note that in practice, we have at most 100,000
nodes in our DAG, so we can achieve a total soundness error of $e$ by setting
the threshold of each individual node at $\frac{1}{10^6}e$.
% TODO
In practice, due to our field size, this error is negligible.

Our definition of zero knowledge is equivalent to the black-box simulator zero knowledge defined in \cite{goldreich1994definitions}.
Since our protocol satisfies this definition, it is also
auxiliary-input zero knowledge, and therefore
has the composability property defined in
\cite{goldreich1994definitions} which shows the full DAG is zero knowledge as well.

\label{bb_protocol}
\subsection{Basic Block Protocols}
We first discuss the relaxation of the algebraic tests for the basic blocks supported in the parallel accumulation protocol. Then we discuss a polynomial interactive oracle proof (Poly-IOP) that the other basic blocks are based on and demonstrate the techniques to prove these special operations.

\subsubsection{Foldable basic blocks}
We support parallel accumulation for 13 basic blocks. Here we introduce the relaxed algebraic test for each basic block. We skip protocols based on cached quotients and refer the readers to Mira \cite{beal2024mira}. We use the notation $f|_\mathbb{H}$ to denote $\{f(\omega^i) \;|\; \forall i\in[n]\}$ and $\omega^n=1$. For convenience, when two commitments $\mathsf{c} \in \mathbb{G}_1$ and $\mathsf{c}' \in \mathbb{G}_2$ appear (possibly with different symbols), we assume they commit to the same underlying polynomial, and therefore $e([1]_1, \mathsf{c}') = e(\mathsf{c}, [1]_2)$ holds.

\minihead{Add, Sub, Eq, and Concat}
We first describe the addition protocol. $\mathcal{P}$ has the private inputs $f(X), g(X), h(X) \in \mathbb{F}_{<n}[X]$, and both parties have access to $\mathsf{f} = [f(\tau)]_1$, $\mathsf{g} = [g(\tau)]_1$, $\mathsf{h} = [h(\tau)]_1$. $\mathcal{P}$ aims to convince $\mathcal{V}$ that $f|_\mathbb{H} + g|_\mathbb{H} = h|_\mathbb{H}$. $\mathcal{P}$ sends $\mathsf{f}, \mathsf{g}, \mathsf{h}$ and $\mathcal{V}$ checks $\mathsf{f} + \mathsf{g} = \mathsf{h}$. In the accumulation scheme, the relaxed verifier check is: $\mathsf{f} + \mathsf{g} - \mathsf{h} = \mathbf{e}_1$. The difference between the vector addition protocol in \cite{beal2024mira} is that we need $g(X)$ to be private for some models (e.g., the residual addition in CNNs). 

The remaining protocols are similar: 1) Sub checks $\mathsf{f}-\mathsf{g}=\mathsf{h}$, 2) Eq checks $\mathsf{f}-\mathsf{g}=0$, and 3) Concat applies Eq on an array of commitments ($\mathsf{f_i}-\mathsf{g_i}=0, \; \forall i$).

\minihead{MulConst, MulScalar, and Mul}
We prove multiplication based on the properties of pairing. Here we first show MulConst: given $\mathcal{P}$ has private inputs $f(X), g(X) \in \mathbb{F}_{<n}[X]$, and both parties have access to their commitments $\mathsf{f}, \mathsf{g}$ and the public constant $c\in\mathbb{F}$. $\mathcal{P}$ aims to convince $\mathcal{V}$ that $c\cdot f|_\mathbb{H} = g|_\mathbb{H}$. $\mathcal{P}$ sends $\mathsf{f}, \mathsf{g}$ and the verifier performs the algebraic test
\[
e(\mathsf{f}, c\cdot[1]_2)=e(\mathsf{g}, [1]_2)
\]
In the accumulation scheme, the relaxed verifier test is:
\[
e(\mathsf{f}, c\cdot[1]_2)-e(\mathsf{g}, [1]_2)=\mathbf{e}_1
\]

Next, we show MulScalar: given $\mathcal{P}$ has the private inputs $f(X), \allowbreak g(X) \in \mathbb{F}_{<n}[X]$ and $S\in\mathbb{F}$, and both parties have access to their commitments $\mathsf{f}, \mathsf{g}, \mathsf{s}\in\mathbb{G}_1$. $\mathcal{P}$ aims to convince $\mathcal{V}$ that $S\cdot f|_\mathbb{H} = g|_\mathbb{H}$. $\mathcal{P}$ sends $\mathsf{f}, \mathsf{g}, \mathsf{s}, \mathsf{s}'$ and the verifier performs the algebraic tests
\[
e(\mathsf{f}, \mathsf{s}')=e(\mathsf{g}, [1]_2), \quad 
e([1]_1, \mathsf{s}')=e(\mathsf{s}, [1]_2)
\]
In the accumulation scheme, the relaxed verifier tests are:
\[
e(\mathsf{f}, \mathsf{s}')-e(\mathsf{g}, [1]_2)=\mathbf{e}_1, \quad
e([1]_1, \mathsf{s}')-e(\mathsf{s}, [1]_2)=\mathbf{e}_2
\]

We then show Mul: given $\mathcal{P}$ has the private inputs $f(X), h(X), \allowbreak g(X) \in \mathbb{F}_{<n}[X]$, and both parties have access to their commitments $\mathsf{f}, \mathsf{h}, \mathsf{g}\in\mathbb{G}_1$. $\mathcal{P}$ aims to convince $\mathcal{V}$ that $f|_\mathbb{H}\cdot h|_\mathbb{H} = g|_\mathbb{H}$. $\mathcal{P}$ computes $t(X)=\frac{f(X)h(X)-g(X)}{X^n-1}$ and its commitment. $\mathcal{P}$ then sends $\mathsf{f}, \mathsf{g}, \mathsf{h}, \mathsf{h}', \mathsf{t}$ and the verifier performs the algebraic tests
\[
e(\mathsf{f}, \mathsf{h}')=e(\mathsf{g}, [1]_2)+e(\mathsf{t}, [\tau^n]_2-[1]_2), \quad 
e([1]_1, \mathsf{h}')=e(\mathsf{h}, [1]_2)
\]
In the accumulation scheme, the relaxed verifier tests are:
\[
e(f, h')-e(\mu\cdot g, [1]_2)-e(\mu\cdot\mathsf{t}, [\tau^n]_2-[1]_2)=\mathbf{e}_1
\]
\[
e([1]_1, \mathsf{h}')-e(\mathsf{h}, [1]_2)=\mathbf{e}_2
\]
where $\mu$ is the slack variable used in the accumulation scheme.

\minihead{Sum, MatMul, and Permute}
We prove these protocols based on the techniques in Aurora \cite{bensasson18aurora}. We first introduce the sum protocol: given that $\mathcal{P}$ has private input $f(X) \in \mathbb{F}_{<n}[X], g\in\mathbb{F}$, and both parties have access to the commitment $\mathsf{f}, \mathsf{g}$, $\mathcal{P}$ proves that $\sum_\mathbb{H} f(X) = g$. $\mathcal{P}$ computes $f_1(X)=\frac{f(X)-f(0)}{X}$ and its commitment, sends $\mathsf{f}, \mathsf{g}, \mathsf{f}_1$ and the verifier performs the algebraic test
\[
e(\mathsf{f}-\mathsf{g}\cdot n^{-1}, [1]_2)=e(\mathsf{f}_1, [\tau]_2)
\]
This is based on the fact that $\sum_\mathbb{H} f(X) = f(0)\cdot n$ \cite{bensasson18aurora}. In the accumulation scheme, the relaxed verifier test is:
\[
e(\mathsf{f}-\mathsf{g}\cdot n^{-1}, [1]_2)-e(\mathsf{f}_1, [\tau]_2)=\mathbf{e}_1
\]

We then show MatMul. Given three private matrices $A, B, C$, where $A$ is an $\ell \times n$ matrix represented by $\ell$ commitments of $f_i(X) \in \mathbb{F}_{\le \ell}[X] \, \forall i \in [n]$, $B$ is an $m \times n $ matrix represented by $m$ commitments of $g_j(X) \in \mathbb{F}_{\le n}[X] \, \forall i \in [m]$, and $C$ is an $\ell \times m$ matrix represented by $\ell$ commitments of $h_j(X) \in \mathbb{F}_{\le m}[X] \, \forall i \in [\ell]$, the MatMul protocol checks that $AB^T = C$. $\mathcal{P}$ and $\mathcal{V}$ sample $\alpha, \beta \in \mathbb{F}$ from randomness $r$ at the start of the protocol, and it is shared over all MatMul and Permute basic block instances. $d_i := \beta^i$, $D(X) := \sum_{i \in [n]} d_iX^i$, and $\mathsf{D'}$ are precomputed by $\mathcal{P}$ and $\mathcal{V}$.

$\mathcal{P}$ computes $a_j := \sum_{i \in [\ell]} \alpha^i A_{ij}$, $A(X) := \sum_{j \in [m]} a_jX^j$, $b_j := \sum_{i \in [n]} \alpha^i B_{ij}$, $B(X) := \sum_{j \in [m]} b_jX^j$, $c_j := \sum_{i \in [\ell]} \alpha^i C_{ij}$, $C(X) := \sum_{j \in [n]} c_jX^j$. $\mathcal{P}$ then computes $l_i := a_ib_i$, $L(X) := \sum_{i \in [m]} l_iX^i$, $Q_l(X) := \frac{A(X)B(X) - L(X)}{X^m-1}$, $L_0(X) := L(0)$, $Q_{l_0}(X) := \frac{L(X) - L(0)}{x}$ and $r_i := \beta^ic_i$, $R(X) := \sum_{i \in [n]} r_iX^i$, $Q_r(X) := \frac{C(X)D(X) - R(X)}{X^m-1}$, $r_0(X) := \frac{1}{m}\sum_{i \in [m]} l_i$, $Q_{r_0}(X) := \frac{r_0(X) - r_0(0)}{x}$.

$\mathcal{P}$ computes and sends $\mathsf{L}, \mathsf{l_0}, \mathsf{A}, \mathsf{B}, \mathsf{B'}, \mathsf{Q_l}, \mathsf{Q_{l_0}}, \mathsf{R}, \mathsf{r}, \mathsf{C}, \mathsf{Q_r}$ and $\mathcal{V}$ performs the tests
\[
e(\mathsf{A}, \mathsf{B'}) - e(\mathsf{L}, [1]_2) - e(\mathsf{Q_l}, [\tau^m-1]_2) = \mathbf{e}_1
\]
\[
e([1]_1, \mathsf{b'})-e(\mathsf{b}, [1]_2)=\mathbf{e}_2, \quad e(\mathsf{L} - \mathsf{l_0}, [1]_2) - e(\mathsf{Q_{l_0}}, [\tau]_2) = \mathbf{e}_3
\]
\[
e(\mathsf{C}, \mathsf{D'}) - e(\mathsf{R}, [1]_2) - e(\mathsf{Q_r}, [\tau^n-1]_2) = \mathbf{e}_4
\]
\[
 e(\mathsf{R} - \mathsf{l_0} \cdot m/n, [1]_2) - e(\mathsf{Q_{r_0}}, [\tau]_2) = \mathbf{e}_5
\]
In the accumulation scheme, the relaxed verifier tests that differ from the original tests are
\begin{align*}
e(\mathsf{A}, \mathsf{B'}) - e(\mu \cdot \mathsf{L}, [1]_2) - e(\mu \cdot \mathsf{Q_l}, [\tau^m-1]_2) &= \mathbf{e}_1 \\
e(\mathsf{C}, \mathsf{D'}) - e(\mu \cdot \mathsf{R}, [1]_2) - e(\mu \cdot \mathsf{Q_r}, [\tau^n-1]_2) &= \mathbf{e}_4
\end{align*}

We finally describe the permute protocol: $\mathcal{P}$ wants to show that
\begin{align*}
&[\alpha^0,\alpha^1,\alpha^2,...] A [\alpha^0,\alpha^n,\alpha^{2n},...]^T\\
= \,&[\alpha^{p_0[0]},\alpha^{p_0[1]},\alpha^{p_0[2]},...] B [\alpha^{p_1[0]},\alpha^{p_1[1]},\alpha^{p_1[2]},...]^T
\end{align*}
where $A$ is a $m \times n$ private input matrix, $B$ is a $m_2 \times n_2$ private output matrix, and $p$ is the permutation.

$\mathcal{P}$ computes $a_j := \sum_{i \in [n]} \alpha^i A_{ij}$, $A(X) := \sum_{j \in [m]} a_jX^j$, $b_j := \sum_{i \in [n_2]} \alpha^{p_0[i]} B_{ij}$, $B(X) := \sum_{j \in [m_2]} b_jX^j$. $\mathcal{P}$ then computes $l_i := \alpha^{n\cdot i}a_i$, $L(X) := \sum_{i \in [m]} l_iX^i$, $c_i := \beta^{n \cdot i}$, $C(X) := \sum_{i \in [m]} c_iX^i$, $Q_l(X) := \frac{A(X)C(X) - L(X)}{X^m-1}$, $l_0(X) := L(0)$, $Q_{l_0}(X) := \frac{L(X) - L(0)}{x}$ and $r_i := \alpha^{p_1[i]}b_i$, $R(X) := \sum_{i \in [m_2]} r_iX^i$, $d_i := \alpha^{p_1[i]}$, $D(X) := \sum_{i \in [n]} d_iX^i$, $Q_r(X) := \frac{B(X)D(X) - R(X)}{X^{m_2}-1}$, $r_0(X) := \frac{1}{m}\sum_{i \in [m]} l_i$, $Q_{r_0}(X) := \frac{r_0(X) - r_0(0)}{x}$.

$\mathcal{P}$ computes and sends $\mathsf{L}, \mathsf{l_0}, \mathsf{A}, \mathsf{B}, \mathsf{Q_l}, \mathsf{Q_{l_0}}, \mathsf{R}, \mathsf{r}, \mathsf{C}, \mathsf{d'}, \mathsf{Q_r}$.
$\mathcal{P}$ and $\mathcal{V}$ compute $\mathsf{C'}, \mathsf{D'}$.
$\mathcal{V}$ performs the tests
\begin{align*}
e(\mathsf{A}, \mathsf{C'}) - e(\mathsf{L}, [1]_2) - e(\mathsf{Q_l}, [\tau^m-1]_2) &= \mathbf{e}_1 \\
e(\mathsf{L} - \mathsf{l_0}, [1]_2) - e(\mathsf{Q_{l_0}}, [\tau]_2) &= \mathbf{e}_2 \\
e(\mathsf{C}, \mathsf{D'}) - e(\mathsf{R}, [1]_2) - e(\mathsf{Q_r}, [\tau^{m_2}-1]_2) &= \mathbf{e}_3 \\
e(\mathsf{R} - \mathsf{l_0} \cdot m/m_2, [1]_2) - e(\mathsf{Q_{r_0}}, [\tau]_2) &= \mathbf{e}_4
\end{align*}
In the accumulation scheme, the relaxed verifier tests that differ from the original tests are
\begin{align*}
e(\mathsf{A}, \mathsf{C'}) - e(\mu \cdot \mathsf{L}, [1]_2) - e(\mu \cdot \mathsf{Q_l}, [\tau^m-1]_2) &= \mathbf{e}_1 \\
e(\mathsf{C}, \mathsf{D'}) - e(\mu \cdot \mathsf{R}, [1]_2) - e(\mu \cdot \mathsf{Q_r}, [\tau^{m_2}-1]_2) &= \mathbf{e}_3
\end{align*}

\subsubsection{Basic blocks based on a Poly-IOP and Div/Mod}\ \\
\label{poly-iop}
The following basic blocks are not accumulated because their proofs contain grand product arguments, creating high-degree constraints, or in the case of Div and Mod they batch the checks for individual division and modulus operations across the tensor into a single check.
We prove Div and Mod with the techniques in \cite{chen2024zkml}. For the other basic blocks, we show how to prove them with a poly-IOP.
Given a multivariate polynomial $R(x_1, x_2, ..., x_{2N})$, the goal of a prover $\mathcal{P}$ is to convince a verifier $\mathcal{V}$ that they know a set of witness polynomials $f_1(x), ..., f_N(x)$ such that $R(f_1(x), ..., f_N(x), \allowbreak f_1(\omega x), ..., f_N(\omega x)) = 0$ for $x=\omega^i, \forall i \in [n]$. Note that $f_i(x)$ is a blinded polynomial in the form of $f_i(x) = g_i(x) + (x^n-1)\cdot b_i(x)$, where $g_i(x)$ is the witness polynomial that contains secret information and $b_i(x)$ is low degree random polynomial. The degree of $b_i(x)$ must be greater than the maximum number of times we evaluate $f_i(x)$ at any particular point, which is $3$ in \sn.

\minihead{Protocol}
 We define groups $\mathbb G_1,\mathbb G_2,\mathbb G_T$, the prime field $\mathbb F$, and a bilinear pairing $e:\mathbb G_1\times\mathbb G_2\rightarrow\mathbb G_T$.
 For $t\in\mathbb F$, let $[t]_s=g_s^t$, where $g_s$ is the generator for group $\mathbb G_s$.
 Since we use the KZG commitment scheme, we assume SDH.
 The protocol proceeds as follows:
\begin{enumerate}[leftmargin=*]
    \item $\mathcal{P}$ sends the commitments of a set of witness polynomials $[f_1(\tau)]_1, ..., [f_N(\tau)]_1$ along with the commitment of a quotient polynomial $[Q(\tau)]_1$ such that $R(f_1(x),..., f_N(x),\allowbreak f_1(\omega x), ..., \allowbreak \\ f_N(\omega x))\allowbreak=Q(x)\cdot (x^n-1)$.
    \item $\mathcal{V}$ uniformly samples $\zeta$ from $\mathbb{F} - \{\omega^i | \forall i \in [n]\}$ and sends $\zeta$.
    \item $\mathcal{P}$ opens all polynomials at $\zeta$ and sends $f_1(\zeta), ..., f_N(\zeta)$, $f_1(\omega\zeta), ..., \allowbreak f_N(\omega\zeta)$ and $Q(\zeta)$.
    \item $\mathcal{V}$ checks if $R(f_1(\zeta), ..., f_N(\zeta)$, $f_1(\omega\zeta), \allowbreak ..., f_N(\omega\zeta))=Q(\zeta)\cdot (\zeta^n-1)$. If so, $\mathcal{V}$ proceeds to ask for the proof of openings; otherwise, $\mathcal{V}$ rejects the proof. $\mathcal{V}$ uniformly samples $\gamma$ from $\mathrm{F}$. Then, $\mathcal{V}$ sends $\gamma$.
    \item $\mathcal{P}$ computes $h_1(x) = \frac{Q(x) - Q(\zeta) + \sum_{k=1}^N \gamma^k [f_k(x) - f_k(\zeta)]}{x-\zeta}$ and \\ $h_2(x) = \frac{\sum_{k=1}^{N} \gamma^{k-1} [f_k(x) - f_k(\omega\zeta)]}{x-\omega\zeta}$. Then, $\mathcal{P}$ sends $[h_1(x)]_1$ and $[h_2(x)]_1$.
    \item $\mathcal{V}$ checks if
    
    $\begin{aligned}[t]
     &e([h_1(\tau)]_1, [\tau-\zeta]_2) \\
    = \,&e\left([Q(\tau)]_1 - [Q(\zeta)]_1 + \sum_{k=1}^N \gamma^k ([f_k(\tau)]_1 - [f_k(\zeta)]_1), [1]_2\right)\\
    \end{aligned}$
    
    and if $\begin{aligned}[t]
     &e([h_2(\tau)]_1, [\tau-\omega\zeta]_2) \\
     = \, &e\left(\sum_{k=0}^{N-1} \gamma^k ([f_k(\tau)]_1 - [f_k(\omega\zeta)]_1), [1]_2\right).\end{aligned}$ 
    
    If so, $\mathcal{V}$ accepts the proof (i.e., output $1$); otherwise, $\mathcal{V}$ rejects it (i.e., output $0$).
\end{enumerate}
The Poly-IOP has perfect completeness, knowledge soundness, and zero knowledge. The proofs are provided in Appendix B.

\minihead{Custom protocols}
We describe four \sn protocols that can be built on top of Poly-IOP in this subsubsection.
\begin{enumerate}[leftmargin=*]
\item \textbf{BooleanCheck}:
The goal of BooleanCheck is to ensure $f(x)$ are either $0$ or $1$ for $x=\omega^i, \forall i \in [n]$. It can be reduced to the Poly-IOP in the following steps: 1) set $N$ as $1$ and let $f_1(x)=f(x)$, and 2) set $R(f_1(x)) = f_1(x)(1-f_1(x))$.

\item \textbf{OrderedCheck}:
Given two degree-\(n\) polynomials \(f(x)\) and \(g(x)\), we want to prove that the sequence \(g(\omega^0), \dots, \allowbreak g(\omega^{n-1})\) is sorted in descending or ascending order with respect to the sequence \(f(\omega^0), \dots, f(\omega^{n-1})\). It can be reduced to three checks.
\begin{itemize}[leftmargin=*]
    \item \textbf{One-to-one Check}: there's a one-to-one mapping between the evaluations of $f(x)$ and $g(x)$. This can be realized by the Permute/CopyConstraint in \sn.
    \item \textbf{Sub Check}: computes and proves that $d(x)=g(\omega x)-g(x)$.
    \item \textbf{Range Check}: proves that every evaluation of $d(x)$ for $x=\omega^i, \forall i \in [n]$ is not greater or less than $0$ if  $g(\omega^0), ..., g(\omega^{n-1})$ is in descending or ascending order, respectively. This can be done with CQ by checking that the evaluations of $d(x)$ are nonnegative or nonpositive.
\end{itemize}

Now, we show that the Sub Check can be reduced to Poly-IOP in the following steps: 1) set $N = 2$ and let $f_1(x)=g(x), f_2(x)=d(x)$, and 2) set $R(f_1(x),f_2(x),f_1(\omega x)) = f_2(x) + f_1(x) - f_1(\omega x)$.

\item \textbf{MaxProof}:
Given two polynomials $f(x)$ and $g(x)$, we want to prove $g(\omega^0)$ is the max element among $f(\omega^0), ..., \allowbreak f(\omega^{n-1})$. It can be reduced to three checks.
\begin{itemize}[leftmargin=*]
    \item \textbf{OrderedCheck}: proves the evaluations of $f'(x)$ are sorted in descending order from $f(x)$. The construction is introduced above.
    \item \textbf{Sub Check}: proves $d(x)=f'(x)-g(x)$. The reduction to Poly-IOP is introduced above.
    \item \textbf{i-thIsZero Check}: proves $d(\omega^i)$ is $0$. In the MaxProof, $i=0$.
\end{itemize}
Denote $L_i(x)$ as a Lagrange polynomial, which is $1$ when $x=\omega^i$ and is $0$ when $x=\omega^j, j\neq i$. We now show \textbf{i-thIsZero Check} can be reduced to Poly-IOP.
\begin{enumerate}[leftmargin=*]
    \item Set $N$ as $2$ and let $f_1(x)=L_i(x), f_2(x)=d(x)$.
    \item Set $R(f_1(x), f_2(x)) = f_1(x) * f_2(x)$.
\end{enumerate}

% \subsubsubsection{SumCheck (from Ari's writeup, not implemented yet)}
% Given two polynomials $s(x)$ and $f(x)$, the goal of SumCheck is to prove $s(\omega^0)=\sum_{i=0}^{n-1} f(\omega^i)$. Note that this is not the famous sumcheck used in GKR protocol.

% First, we construct a polynomial $z(x)$ such that $z(\omega x)-z(x) = f(x)$ and $z(\omega^0)=0$. By the construction, $z(\omega^n)$ is the sum we aims to prove.

% SumCheck can be reduced to three PolyChecks: $z(\omega x)-z(x) - f(x) = 0$, $L_0(x)z(x)=0$, and $L_n(x)[z(x)-s(x)]=0$ by the above techniques.

\item \textbf{CopyConstraint}:
This basic block is based on the permutation argument in Plonk \cite{gabizon2019plonk}, additionally supporting input and output polynomials to have different degrees and padded values without needing to construct new input and output commitments. The construction is as follows:

Let $H \subset \mathbb{F}$ be a multiplicative subgroup of order $N = \max(m, n)$.
  We define $L_i(X)$ to be such that ${L_i}$ for $i \in [\max(m, n)]$ is a Lagrange basis for $H$.

  Input polynomials: $f_i(X) = \sum_{j=1}^m w_{i\cdot m + j}L_{N/m \cdot j}(X)$ for $i \in [p_1]$
  
  Output polynomials: $g_j(X) = \sum_{k=1}^n w_{j\cdot n + p_1 m + k}L_{N/n \cdot k}(X)$ for $j \in [p_2]$ for a witness $(w_i)_{i\in[p_1m + p_2n]}$

  Define $(f_{(1)}, \ldots, f_{(p_1m)}) \in \mathbb{F}^{p_1m}, (g_{(1)}, \ldots, g_{(p_2n)}) \in \mathbb{F}^{p_2n}$ by
  $f_{((j_1 -1) m + i_1)} := f_{j_1}(\omega^{i_1}), g_{((j_2 - 1) n + i_2)} := g_{j_2}(\omega^{i_2})$ for $j_1 \in [p_1], i_1 \in [m], j_2 \in [p_2], i_2 \in [n]$
  
  CopyConstraint ensures that for a set of padding partition indices \textsf{pad\_idx}, mapping \textsf{pads}: \textsf{pad\_idx} $\rightarrow$ \textsf{pad\_val} and $\sigma: [p_2n] \rightarrow [p_1m] \cup \textsf{pad\_idx}$,
  \[   
g_{(\ell)} = 
     \begin{cases}
        \textsf{pads} (\sigma(\ell)) & \sigma(\ell) \in \textsf{pad\_idx} \\
       f_{(\sigma(\ell))} & \sigma(\ell) \not\in \textsf{pad\_idx} \\
     \end{cases}
\]
  The partition construction $\sigma$ is constructed in the same way as in Plonk \cite{gabizon2019plonk}. However, unlike the permutation argument in Plonk, we additionally support cases where the input and output degree are different. To handle padding, all padded indices with the same pad value are contained in the same partition with indices in \textsf{pad\_idx}. Let $q$ be the number of pad values, and $\textsf{pad}_i$ be the $i$-th pad value. For each $i \in [q]$, let $r_i$ be the smallest value such that $\textsf{pad}_i$ is constrained to be a coefficient of $G_i(X) = f_{r_i}(X)$ or $G_i(X) = g_{r_i-p_1}(X)$. Let $t_i$ be the smallest value such that $G_i(X)L_{t_i}(X) = \textsf{pad}_i$.
  If $m < N$ or $n < N$, when $i\mod \frac{N}{m} \neq 0$ or $i\mod \frac{N}{n} \neq 0$, then the element will be in its own singleton partition.

CopyConstraint reduces to Plonk's Section 5 extended copy constraints \cite{gabizon2019plonk}. It is sufficient to call the extended copy constraints as a subroutine with the input and output polynomials and $\sigma$ described above and perform the following Poly-IOP for padding: for $i \in [q]$, $G_i(x)L_{t_i}(x) = \textsf{pad}_i$.

Overall, CopyConstraint reduces to these Poly-IOPs:

\begin{itemize}[leftmargin=*]
\item \textbf{One Check}: $L_1(x)(z(x) - 1) = 0$

\item \textbf{Permutation Check}: 
    
    $\begin{aligned}[t] &z(x)\prod_{i=1}^{p_1} (f_i(x) + i\beta x + \gamma)\prod_{i=1}^{p_2} (g_i(x) + i\beta x + \gamma) \\
= \, &z(\omega x)\prod_{i=1}^{p_1} (f_i(x) + \beta S_{\sigma_i}(x) + \gamma)\prod_{i=1}^{p_2} (g_i(x) + \beta S_{\sigma_{p_1 + i}}(x) + \gamma)
\end{aligned}$

\item \textbf{Pad Check}: For $i \in [q]$, $G_i(x)L_{t_i}(x) = \textsf{pad}_i$, where

$S_{\sigma_i}(X) = \sum^{m}_{j=1} \sigma(i \cdot m +j)L_{N/m \cdot j}(X)$ for $i \in [p_1]$
  
  $S_{\sigma_{p_1+i}}(X) = \sum^{n}_{j=1} \sigma(i \cdot n + p_1m +j)L_{N/n \cdot j}(X)$ for $i \in [p_2]$

  $\begin{aligned}[t]
 z(X) = &\sum_{i=1}^{N-1} \left( L_{i+1}(X) \prod_{j=1}^i \prod_{k=1}^{p_1+p_2} \frac{(w_{kN + j} +\beta k\omega^j+\gamma)}{(w_{kN + j} +\sigma(kN+j)\beta+\gamma)} \right) \\
 &+ L_1(X)
\end{aligned}$
\end{itemize}
\end{enumerate}

\section{Compiler and Transpiler}

The remaining key components in \sn are the compiler and transpiler. With these components, \sn can transform a given ML model into the underlying basic blocks. Since the prover efficiency of different basic blocks varies and \sn does not support accumulation for some special basic blocks, we need to carefully design the transformation rules to keep \sn performant in terms of the prover overhead and the proof size. There are two major challenges in constructing such a compiler.

First, we must optimize within each layer for performance while maintaining high accuracy. For example, when performing shape operations, we opt for the cheapest sequence of operations.

For some basic blocks, we choose to prioritize accuracy, such as softmax. The softmax function over a vector of size $K$ elementwise is $\sigma(\mathbf{z})_i = e^{z_i} / \sum_{j=1}^K e^{z_j}$. This can naively be implemented with an exponentiation, sum, and division. However, the elements of the input $\mathbf{z}$ can result in exponentiation exceeding the lookup table range. Instead, we compute $m = \max(\mathbf{z})$, and the exponentiation $e^{z_i-m}$ which will be a small value since the exponent is negative. 

Second, we must optimize \emph{across} layers for high performance. For this, we create graph transformation rules that prune the DAG of basic blocks and combine basic blocks as well.

% Third, we must choose scale factors to achieve high accuracy while maintaining
% reasonable table sizes. Choosing a uniform scale factor that is too large
% results in tables that are too large to prove (e.g., table sizes of $2^{50}$)
% or poor accuracy.
% Therefore we use different scale factors at different points in the computa

We support all the MLPerf Inference: Edge V4.1 models \cite{mlperf}, which include state-of-the-art LLMs, CNNs, and RNNs. We now provide an overview of \sn's compiler before discussing its implementation.

\subsection{Overview}
\sn consists of three steps: 1) graph compilation to optimize ML layers for cryptographic proving as opposed to standard inference, 2) transpilation of the compiled ML models to a DAG composed of basic blocks, and 3) proving and accumulation on the resulting DAG.

Speicifically, \sn's compiler takes as input an ML model specification (we use ONNX for our implementation) and outputs an optimized, compiled ML model (another optimized ONNX) that will be used in the downstream task. 

The transpiler then takes the optimized ML model specification and
outputs a program specific to the model at hand. Furthermore, several parts of the proving process can be
precomputed per-model to reduce the overall proving time per inference.

In our current implementation, we use rule-based graph transformations for the model
compilation and transpilation.
% We run the model through a sample input to select the scale factors used at different points in the computation.
We now describe the different transformation rules in \sn compiler.

\subsection{Graph Transformation Rules}
Graph transformations (compilation) are a widely used method to improve ML model performance.
The same operation can be computed in many different ways, which have different
performance characteristics on different hardware platforms for standard ML
inference. Similarly, the performance characteristics of identical methods of
computation widely vary for our basic blocks.

Our graph transformation implementation takes as input an ONNX file and outputs
an ONNX file optimized for \sn. We use rule-based transformations to use fewer or
cheaper cryptographic operations. In this work, we implement seven
transformation rules and apply them greedily until the graph does not change.
This optimization method is commonly used in compilers
\cite{alfred2007compilers}.

\begin{table*}[t]

\centering

\caption{List of graph transformation rules we currently implement in \sn. Figure \ref{fig:bb1}a shows GeLU.}
\label{table:rules}

\begin{tabular}{l|p{2.1cm}|p{3cm}|p{7.4cm}}

Rule name      & Model                     & Reason         & Description                                                                                                                                                                      \\
\hline
ReshapeTrans    & Bert                      & Decreases number of permutes from 2 to 1 &
  Replace every Reshape node followed by a Transpose node with a custom ReshapeTrans node. \\
  [1ex]

GeLU            & GPTj                      & Decreases number of CQs from 6 to 1 &
  Replace every subgraph that represents GeLU with one GeLU node. \\
  [1ex]

MultiHeadMatMul & GPTj                      & Uses no copy constraint &
  Replace every subgraph that represents MultiHead MatMul with one custom MultiHeadMatMul node. \\
  [1ex]

RoPE            & GPTj                      & Uses no copy constraint &
  Replace every subgraph that represents RoPE operation with two custom nodes (RoPEConst and RoPERotate). \\
  [1ex]

CustomCNN       & ResNet, 3DUnet, RetinaNet & Uses no copy constraint, cheap aggregation  &
  Execute convolutions by viewing intermediate tensors as channels second. This allows us to use only cqlin and add.  \\
  [1ex]

MultiHeadConv   & RetinaNet                 & Uses no copy constraint               &
  Replace conv followed by reshape and transpose with a custom MultiHeadConv. \\ 
  [1ex]

ConcatConv      & 3DUnet                    & Uses no copy constraint               &
  Replace two conv followed by a concat with a custom ConcatConv to avoid the concat. \\

\end{tabular}

\end{table*}

\begin{figure}[t!]
\minipage{0.45\textwidth}
  \includegraphics[width=\textwidth]{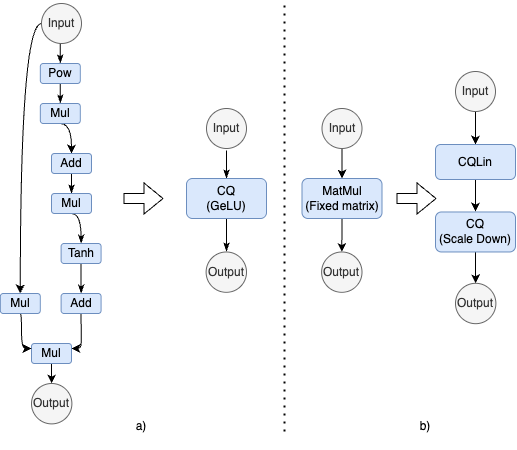}
  \caption{a) One CQ for GeLU, b) CQLin and CQ for MatMul with one fixed matrix }
  \label{fig:bb1}
\endminipage\hfill
\end{figure}

We provide a summary of the rules we use in Table~\ref{table:rules}. Our graph transformation rules do not change the overall model architecture.
They instead replace common layer patterns in ONNX files with an optimized implementation.
For example, the GeLU operation in an ONNX file is expressed as 8 different arithmetic layers including Pow and Tanh. We can replace these layers with 
a single GeLU layer that is implemented with a single table lookup that directly performs the GeLU mapping. This transformation is shown in Figure \ref{fig:bb1}. Our transformation rules can provide up to a $10\times$ 
speedup for the operations at hand and up to a $4.8\times$ speedup for real-world models. Moreover, \sn's graph transformation rules are easily extensible and can be updated to accommodate newer models.

% \subsection{Scale Factor Selection}

% ML models are usually run on floating point numbers, which have a high overhead if proved in zero knowledge.
% Therefore it is necessary to quantize our model to use fixed point numbers instead.
% In fixed point, a number $a$ is stored as $\lfloor a\cdot\texttt{SF}\rceil$ where $\texttt{SF}$ is the scale factor.
% We have found that using a fixed scale factor across the entire model computation often doesn't work with the model accuracy and available memory requirements we have.
% The issue is that if the scale factor is too small, we lose too much accuracy, and if the scale factor is too large, we run out of memory because we have to store very large lookup tables, as the fixed point tensors become very large.

% Therefore we use a different scale factor at different points in the computation. Intuitively, when a tensor in the computation has larger numbers in it, it should have a larger scale factor.
% In order to calculate the scale factors, we run the ML model through a sample input, and measure the value of the tensor as is passes through the model.
% Then we use these values to calculate the scale factors to use.

% \todo{}{this section is very handwavy, need to figure out how to expand it or reduce its importance in the rest of the paper}

\subsection{Transpilation}
Given an optimized graph, \sn's final step is to
transpile the graph into an executable cryptographic protocol. Similar to our
graph transformation rules, we currently implement our transpiler with
rule-based transformations.

We implement 61 layers in the transpiler, corresponding to ONNX layers. Due to space
limitations, we cannot fully describe all of the layer implementations here.
Instead, we describe classes of layers and highlight specific instances for
illustrative purposes.

Before describing the classes of layers, we highlight several common patterns.
First, any layer that multiplies two numbers must have a rescaling at some point
after that layer to correct for the scale factor \cite{chen2024zkml}. Second, due to the way we lay
out the committed intermediate values, which we call \emph{activations}, certain operations can be
free. Our graph transformations aim to lay out the activations so that they use
as few expensive shape operations as possible.

\minihead{MatMul-based layers}
The Gemm, MatMul, LSTM, Conv, and ConvTranspose layers are implemented with matrix multiplication operations. When one matrix is known at compile time, such as the weights in Conv or with a fixed MatMul matrix, we prove matrix multiplication using CQLin. Figure \ref{fig:bb1}b shows how a MatMul layer is transformed into basic blocks when one matrix is fixed.

\minihead{Shape operations}
ML models frequently employ layer operations that manipulate array shapes, including Reshape, Concat, Split, and Transpose operations. For both input and output tensors, polynomial commitments are made along the tensor's final axis. When shape operations preserve this final axis, no new polynomial commitments are required for the outputs. However, many such operations still modify the final axis. In this case, transpose and most reshape operations can be proven using our Aurora-based Permute protocol. For more specialized cases involving element copying, we implement CopyConstraint, which offers the most flexibility but comes with higher computational costs. For example, we must still use CopyConstraint in the Slice layer, where we want to copy a subset of consecutive elements into a new vector.

\minihead{Pointwise non-linearities}
We use the CQ2 basic block to verify nonlinearities and scale factor corrections after multiplication and division operations. Whenever possible, we combine nonlinearities and scaling operations within a layer to reduce the number of required lookup proofs. Since CQ2 is accumulable, the number of nonlinearities in an ML model does not affect performance, which means that replacing any nonlinearity with another in \sn does not increase the prover overhead or proof size.
% ML models typically contain nonlinearities to capture non-linear relationships and decision boundaries. We also need to introduce scaling operations to correct for scale factors changing after multiplication and division operations. Both of these operations are proven by checking for inclusion in a lookup table using the CQ protocol \cite{eagen2022cq}. When possible, we combine nonlinearities and scaling operations within a layer to reduce the number of lookup operations to prove.

\minihead{Arithmetic operations}
Most arithmetic layers such as Add, Mul are trivially implemented using a corresponding basic block. We describe two examples using custom basic blocks:

\textbf{And}: To implement And, we use BooleanCheck to check that each input array only contains 0s or 1s, and then use Mul to obtain their product.

\textbf{ArgMax}: To implement ArgMax, we use OrderedCheck to check that the sorted input is in descending order, One-to-oneCheck to check that the sorted elements correspond to the original unsorted input, and CopyConstraint to select the first element.

\minihead{Bespoke operations}
To efficiently prove ML models, \sn replaces some specific patterns of ONNX nodes with customized operations. This is because if \sn directly encodes and proves based on the original tensors, the proving overhead will be too expensive for practical usage. Here we describe two examples of bespoke operations:

\textbf{CustomConv}: We can replace all the Conv with CustomConv in CNNs, achieving the same accuracy as Conv, while saving memory, proving time, and proof size by removing the need for any shape operation in Convolution. For example, in a 2D convolution, rather than feeding in an input of size $[B, I, H,W]$, \sn makes all input sizes become $[BHW, I]$. In other words, all the original inputs $X$ in our CustomCNN are viewed as $\tilde{X}$:
\[
\tilde{X} \in \mathbb{F}^{(BHW) \times I}, 
\quad
\tilde{X}\bigl[b \cdot (HW) + r \cdot W + c,\, i\bigr] 
\;=\;
X[b,\,i,\,r,\,c].
\]

Let the dimensions of kernel $W$ be $[O, I, h, w]$. \sn can transpile the channel update as $hw$ matrix multiplications (CQLins) between the input $\tilde{X}$ and a known 2D matrix $\tilde{W}_{p,q}$ of shape $[O, I]$:
\[
\tilde{W}_{p,q} \in \mathbb{F}^{OI},
\quad
\tilde{W}_{p,q}[\,o,\,i\,] 
\;=\;
W[o,\,i,\,p,\,q].
\]

Each CQLin will then result in a tensor $\tilde{Y}_{p,q}$ of shape $[BHW, O]$. 

\[
\tilde{Y}_{p,q}
\;=\;CQLin(
\tilde{X}
\,,\,
\bigl(\tilde{W}_{p,q}\bigr)^T)
\;\;\in\;
\mathbb{F}^{(BHW) \times O}.
\]

Later, we select some of the tensors based on the window sliding mechanism of Conv and add them together. For simplicity, we consider Conv without padding. Let $H', W'$ be the new height and width after convolution in the original Conv layer. We can compute $Y'\in\mathbb{F}^{(BH'W')\times O}$, where $Y\bigl[b \cdot (H'W') + r \cdot W' + c,\, o\bigr]$ equals:
\[
\sum_{p=0}^{h-1}
\;\sum_{q=0}^{w-1}
\tilde{Y}_{p,q}\Bigl[
b \cdot (HW) 
\;+\; (r + p) \cdot W 
\;+\; (c + q),
\; o\Bigr].
\]
The value selection is free because we only encode the last dimension into commitment and the addition is also almost free because of the homomorphism of the KZG commitment. In other words, the $b \cdot (H'W') + r \cdot W' + c$ elements in $[Y]_1\in\mathbb{G}^{BH'W'}$ can be computed with $[\tilde{Y}_{p,q}]_1\in\mathbb{G}^{BHW}$ during the proving phase by: 
\[
\sum_{p=0}^{h-1}
\;\sum_{q=0}^{w-1}
[\tilde{Y}_{p,q}]_1\Bigl[
b \cdot (HW) 
\;+\; (r + p) \cdot W 
\;+\; (c + q)\Bigr]
\]

$Y$ can then directly be used as the input of the next CustomConv. To understand what $Y$ is, we denote the output after the original Conv as $Y'\in\mathbb{F}^{B\times O\times H'\times W'}$, we can see:
\[
Y\bigl[b \cdot (H'W') + r \cdot W' + c,\, o\bigr] 
\;=\;
Y'[b,\,o,\,r,\,c].
\]

For example, suppose we want to perform a 2D convolution on an image with a tensor of shape $C \times D \times D$. We first need $32 \times 32$ CQLins to change the number of channels from $3$ to $C$. We perform the almost-free additions introduced in Section 5.3 to generate the $D \times D$ tensor. We finally need $D \times D$ CQs to scale the output back. We can fold the proofs into only one CQLin, one CQ, and one addition in the end with generalized Mira. 
    
\textbf{GeLU}: As shown in Fig. \ref{fig:bb1}, there are 4 Mul, 1 Pow, and 1 Tanh in the original ONNX GeLU activation function (approximated by $0.5x \left( 1 + \tanh\left[ \sqrt{\frac{2}{\pi}} \left( x + 0.044715x^3 \right) \right] \right)
$). To prevent using multiple CQ and Mul basic blocks, we combine all operations into one CQ.
\section{Experiments}
We now turn to evaluating \sn. We show that \sn can prove inference on a wider
range of model than prior work: it can prove all of the MLPerf edge inference
models on a single server. 
% We further show that \sn's compiler optimizations
% result in high performance gains of up to $38 \times$. This was calculated based on our estimate of how long the ZKML framework \cite{chen2024zkml} would take to prove LLaMA-2-7B, versus the proving numbers we measured in \sn.

\subsection{Experimental Setup}
\label{sec:setup}
\begin{table}[t]
\caption{List of models considered in the evaluation with their parameters and input dimensions.}
    \centering
    \begin{tabular}{ll|ll}
     Model & Part & Parameters & Input dimensions  \\
    \hline
  BERT      & & 110M  & 1 token  \\
  GPT-j  & & 6B  & 2 tokens  \\
  LLaMA-2-7B    & & 7B    &  1 token \\
  RNNT &  & 120M     & seqlen=1, batch=1  \\ 
  SDXL & Text encoder & 3.5B   & 1 token \\
  & Text encoder 2 &  & 1 token \\
  & VAE decoder &   & 1x4x4x4  \\
  & UNet & & 1x4x4x4, seqlen=1 \\
  ResNet-50  & & 25.6M  & 1x3x224x224 \\
  RetinaNet  & & 34M   & 1x3x800x800  \\
  3D-UNet    & & 19M     & 1x1x128x128x128 \\
    \end{tabular}
    \label{table:params}
\end{table}
We evaluate proving time, verification time, proof size, and accuracy on all models in the MLPerf Inference v4.1 Edge benchmark, as well as RNNT from v4.0 and LLaMA-2-7B. These include ResNet50-v1.5, RetinaNet 800x800, BERT, 3D-Unet, GPT-j, and Stable-Diffusion-XL. The number of parameters and input sizes we used for each model is shown in Table \ref{table:params}. Our experiments are all conducted on a pairing-friendly elliptic curve BN254 over a 254-bit prime field. For all models we use 10 bits of quantization except for GPT-j, where we used 12 bits.

Proving and verification performance are measured on a single server with an Intel(R) Xeon(R) Platinum 8358 2.6 GHz processor, with 64 threads and 4 TB memory.

\subsection{Comprehensive End-to-End Results on the Entire MLPerf Inference Suite v4.1}
\begin{table*}[t]
\centering
\caption{End-to-end proving time for MLPerf Inference v4.1 Edge models, RNNT, and LLaMA-V2-7B.}
\begin{tabular}{llrrrr}
\toprule
\textbf{Model} & \textbf{Part} & \textbf{Proving time} & \textbf{Verification time} & \textbf{Proof size} \\
\midrule

BERT
  & \textit{(N/A)}    & 880.42 s   & 26.67 s    & 4.88 MB \\
\cline{1-5}

GPT-j
  & \textit{(N/A)}    & 1397.52 s  & 62.64 s    & 6.54 MB \\
\cline{1-5}

LLaMA-2-7B
  & \textit{(N/A)}    & 2645.50 s  & 100.14 s   & 22.85 MB \\
\cline{1-5}

RNNT
  & Joint        & 1.33 s    & 0.02 s    & 11.13 KB \\
  & Encoder      & 15.94 s   & 0.85 s    & 31.23 KB \\
  & Prediction   & 2.12 s    & 0.19 s    & 17.10 KB \\
  & \textbf{Total} & 19.39 s   & 1.06 s    & 59.46 KB \\
\cline{1-5}

SDXL
  & Text encoder   & 505.64 s  & 6.34 s   & 249.7 KB \\
  & Text encoder 2 & 742.43 s  & 21.42 s  & 986.54 KB \\
  & VAE decoder    & 23575 s   & 391.81 s & 71.74 MB \\
  & UNet           & 44127 s   & 583.68 s & 72.86 MB \\
  & \textbf{Total} & 68950 s   & 1003.3 s & 145.84 MB \\
\cline{1-5}

ResNet-50
  & \textit{(N/A)}   & 6270.50 s & 62.81 s  & 85.41 KB \\
\cline{1-5}

RetinaNet
  & \textit{(N/A)}   & 53052 s   & 1171.3 s & 98.7 KB \\
\cline{1-5}

3D-UNet
  & \textit{(N/A)}   & 568543 s  & 14406 s  & 1.4 MB \\

\end{tabular}
\label{table:e2e}
\end{table*}

We now present our end-to-end proving results, demonstrating \sn as the first zero-knowledge system capable of handling \emph{all} models in the MLPerf Inference suite v4.1. Table~\ref{table:e2e} reports proving time, verification time, and proof size across a diverse range of architectures, showcasing how \sn seamlessly scales from smaller CNNs to large-scale transformers within a unified ZK framework. These results illustrate the practical performance implications, namely, the computational overhead and lightweight proofs, of applying \sn to modern deep learning inference tasks.

In addition, we use the similar scaling techniques proposed in ZKML \cite{chen2024zkml}. Table~\ref{table:acc} shows that \sn's quantization and choice of scale factors preserve inference accuracy (e.g., ROUGE and mAP) at the same level as standard MLPerf baselines, which validates the functional correctness of \sn by comparing accuracy metrics (e.g., ROUGE, F1, WER, mAP) against standard MLPerf baselines, showing that zk-compatible execution does not degrade model quality.
\begin{table}[t]

\centering

\caption{Accuracy numbers with \sn scaling.}

\begin{tabular}{l|l|l|l}
Model     & Metric   & \sn Accuracy                                                                       & Ref. Accuracy                                                                                     \\\hline
GPT-j      & ROUGE    & \begin{tabular}[c]{@{}l@{}}ROUGE 1 - 43.4672\\ ROUGE 2 - 20.2609\\ ROUGE L - 30.0066\end{tabular} & \begin{tabular}[c]{@{}l@{}}ROUGE 1 - 42.9865\\ ROUGE 2 - 20.1235\\ ROUGE L - 29.9881\end{tabular} \\
BERT      & F1 score & 90.104\%                                                                                          & 90.874\%                                                                                          \\
ResNet-50    & Accuracy & 76.038\%                                                                                          & 76.456\%                                                                                          \\
RNNT      & WER      & 7.49\%                                                                                            & 7.45\%                                                                                            \\
3D-UNet    & Accuracy & \begin{tabular}[c]{@{}l@{}}Mean - 86.18\%\\ Kidney - 93.47\%\\ Tumor - 78.88\%\end{tabular}       & \begin{tabular}[c]{@{}l@{}}Mean - 86.17\%\\ Kidney - 93.47\%\\ Tumor - 78.87\%\end{tabular}       \\
RetinaNet & mAP      & 0.3753                                                                                            & 0.3757                                                                          
\end{tabular}
\label{table:acc}
\end{table}

\subsection{Comparisons with Prior Work}
We compare the performance of \sn to the general-purpose framework~ZKML \cite{chen2024zkml} by benchmarking the most expensive model GPT-2 in ZKML. \sn achieves a proving time of 599s, representing a 6$\times$ speedup (ZKML proves GPT-2 in 3601s within the identical hardware environment). Furthermore, the verification time is reduced to 12s, outperforming the 19s reported by ZKML. Another general proving is ezkl \cite{ezkl}, which we do not compare against since zkml is faster than ezkl due to its circuit optimizer. Concretely, ezkl on nanoGPT small (1-million parameters, 26M flops) needs 966s to prove, while zkml on a larger LLM GPT-2 (117-million parameters, 189M flops) requires only 159s to prove per token. Considering this, we only used zkml in the evaluation to compare the prover performance. 

When comparing proof sizes, we find that \sn even outperforms specialized protocols including zkCNN \cite{liu2021zkcnn}, Mystique \cite{weng2021mystique}, and pvCNN \cite{weng2022pvcnn} by at least $3\times$ due to the adoption of the accumulation scheme, as shown in Table \ref{table:proof_size}. Following Mira \cite{beal2024mira}, we do not compare the performance of zkLLM\cite{sun2024zkllm} with \sn for the same reason: zkLLM is GPU-based proving framework, whereas \sn serves for high-performance CPU-intensive environments.
\begin{table}[t]
\centering
\caption{Comparisons of proof size. (*Did not implement ResNet-50, we use the proof size of much smaller models (i.e., VGG-16 for zkCNN and LeNet-5 for pvCNN.))}
\begin{tabular}{l|c|c|c|c}
Framework & zkCNN\cite{liu2021zkcnn} & Mystique\cite{weng2021mystique} & pvCNN\cite{weng2022pvcnn} & \sn \\\hline
ResNet-50               & > 304 KB*             & 1.27 GB             & > 343 MB*                & 85.41 KB                  
\end{tabular}
\label{table:proof_size}
\end{table}

\subsection{Ablation Studies}
To understand the power of the parallel accumulation scheme in \sn, we compare the runtimes of the prover and verifier  between \sn and the sequential accumulation based on Mira~\cite{miranda2021towards} across five representative models, as shown in Table~\ref{table:mira_onnx_combined}. ZkTorch consistently outperforms Mira in both proving and verification phases, demonstrating significant runtime improvements. For example, in the case of GPT-j, the proving time is reduced from 8663s to 1398s, a $6.2\times$ speedup, while the verifier time drops from 239s to 63s. BERT sees a $1.3\times$ speedup in both proving and verification. The most substantial gain is observed in ResNet-50, where proving time is halved and verification time improves by over $16\times$. Similar trends are observed for LLaMA-2 and RNNT. These results highlight the optimization of accumulation protocol introduced in \sn, which reduces end-to-end overhead while preserving correctness guarantees and the same size of the resulting proof.

To evaluate the effectiveness of the \sn compiler, we compare the proving time, verification time, and proof size before and after compilation across three representative models: GPT-j, BERT, and ResNet-50. As shown in Table~\ref{table:proof_size}, ZkTorch yields substantial reductions across all metrics. For GPT-j, proving time is reduced from 5181s to 1398s (a $3.7\times$ speedup), while verification time drops from 383s to 63s. Similarly, BERT demonstrates a dramatic improvement, with proving time decreasing from 52977s to 880s. Notably, proof size is also significantly compressed from 155 MB to just 4.88 MB. ResNet-50 shows consistent improvements, with a $4.8\times$ reduction in proving time and a $2,398\times$ reduction in proof size. These results highlight the compiler's effectiveness in optimizing zk-SNARK execution pipelines, reducing both computational and communication overhead while maintaining functional correctness.

\begin{table}[t]
\centering
\caption{Prover and verifier time with and without parallelized Mira. Para. is to denote parallelized or not, and $t_\mathcal{P}$ and $t_\mathcal{V}$ are the prover and verifier time}
\begin{tabular}{ll|c|c|c|c|c}
\toprule
 & Para. & GPT-j & BERT & ResNet-50 & LLaMA-2 & RNNT \\
\midrule
$t_\mathcal{P}$
 & Off & 8662.96s & 1118.25s & 12312.87s & 5976.34s & 35.62s \\
 & On & 1397.52s & 880.42s & 6270.51s & 2645.50s & 19.39s \\
\midrule
$t_\mathcal{V}$
 & Off & 239.32s & 35.99s & 1013.22s & 206.06s & 1.76s \\
 & On & 62.64s & 26.67s & 62.81s & 100.14s & 1.06s \\
\bottomrule
\end{tabular}
\label{table:mira_onnx_combined}
\end{table}

\begin{table}[t]
\centering
\caption{Prover, verifier time, and proof size with and without the \sn compiler}
\begin{tabular}{ll|c|c|c}
\toprule
 & Compiler & GPT-j & BERT & ResNet-50 \\
\midrule
Prover time
 & Off  & 5181.47s   & 52976.51s  & 30351.17s \\
 & On & 1397.52s   & 880.42s    & 6270.51s  \\
\midrule
Verifier time
 & Off  & 382.85s    & 979.87s    & 735.90s   \\
 & On & 62.64s     & 26.67s     & 62.81s    \\
\midrule
Proof size
 & Off  & 115.39 MB  & 154.87 MB  & 200.01 MB \\
 & On & 6.54 MB    & 4.88 MB    & 85.41 KB  \\
\bottomrule
\end{tabular}
\label{table:compiler_prover}
\end{table}
% \subsection{Detailed Analysis}

% TODO

\section{Related Work}

\minihead{ZK-SNARK protocols}
ZK-SNARKs have been widely studied in the security community. These protocols
include protocols with different properties. For example, Groth16 has constant
verification time with a per-circuit setup \cite{groth2016size}, while Spartan has linear
proving time but longer proofs \cite{setty19spartan}. These protocols have also been
specialized for other applications, such as general-purpose ZK virtual machines
\cite{succinct} and ML, which we discuss below. Our work utilizes KZG vector commitments \cite{kate2010constant} and draws inspiration from
prior protocols, including Aurora \cite{bensasson18aurora}.

\minihead{ZK-SNARKs for ML}
Researchers have developed ZK-SNARK protocols to perform ML inference. These
protocols broadly fall under two categories.

The first category are compilers that take the computation in ML inference and
compile them to generic ZK-SNARK protocols. These include ezkl, zkml, and others
\cite{ezkl,chen2024zkml}. These compilers compile layers into a circuit in the halo2 framwork, a general ZK-SNARK framework, which is based on Plonk, a general ZK-SNARK proving system. Because generic ZK-SNARK protocols are universal, these systems can
support a wide range of ML models. However, they suffer in performance compared
to specialized systems. Instead, the basic blocks in \sn contain custom protocols for different common operations and \sn compiles layers into a graph of basic blocks.

The second category are specialized protocols for specific kinds of models.
These include specialized protocols for CNNs \cite{liu2021zkcnn} and LLMs \cite{sun2024zkllm}.
While they are more efficient than generic ZK-SNARK protocols, they do not
support a wider range of models. Furthermore, \sn is the first general ZKML framework that supports accumulation scheme to our best knowledge. This allows \sn to produce proofs that are even much smaller than specialized protocols such as zkCNN.
% Furthermore, as we describe in
% Section~\ref{sec:issues}, they can be insecure or result in vacuous accuracy.
%\todo{bing-jyue}{compare zkcnn, mystique, zkllm}

\vspace{0.5em}

In this work, we develop a specialized compiler that generates custom protocols
for AI and ML models, bridging the gap between generic ZK-SNARK protocols and
specialized protocols for specific model types. %To the best of our knowledge, our work is the first to do so in a high-performance manner.

\minihead{Accumulation for ZK-SNARKs}
In \sn, we use accumulation to make the proof size of repeated basic blocks constant size. We adapt Mira \cite{beal2024mira}, a accumulation scheme designed for pairing-based arguments. Mira verification is inherently sequential and the work only discusses how to support few protocols (e.g., CQ and CQLin) with the scheme. We extend Mira for improved parallelism, and also apply this scheme to 10 more basic blocks.
\section{Conclusion}

We present \sn, a general-purpose framework for compiling machine learning models into zero-knowledge proofs. \sn decomposes high-level models into a directed acyclic graph of basic cryptographic blocks, enabling support for a wide range of architectures including CNNs, RNNs, and transformers. In total, \sn supports 61 commonly used ML layers. To improve scalability, \sn extends the Mira accumulation scheme with a parallelized aggregation protocol, allowing efficient composition of many proof instances and significantly reducing overall proving time. This parallel accumulation mechanism plays a central role in making proof generation feasible for large-scale models. \sn further incorporates a compiler that optimizes the proof graph by applying structure-aware transformations. Finally, our experiments across diverse ML models shows the benefits \sn achieves with respect to speed, accuracy, and generalizability.
\section*{Acknowledgements}
We would like to acknowledge the Open Philanthropy project and the VAIL project for funding this research in part.

\balance
\bibliographystyle{ACM-Reference-Format}
\bibliography{paper}
\clearpage

\renewcommand{\thesection}{A}
\section{Code}

Our open-source implementation is available at:\\
\url{https://anonymous.4open.science/r/zkt-2477}
\renewcommand{\thesection}{B}
\section{Security properties of Poly-IOP}
We use the notation in Sec \ref{poly-iop} to prove the security properties:
\begin{enumerate}[leftmargin=*]
    \item \textbf{Completeness}
    If $\mathcal{P}$ has a set of witness polynomials $f_1(x), ..., \\ f_N(x)$ such that $R(f_1(x), ..., f_N(x), \allowbreak f_1(\omega\zeta), ..., \allowbreak f_N(\omega\zeta)) = 0$ for $x=\omega^i, \forall i \in[n]$, the proof will be accepted with probability $1$.

    \item \textbf{Knowledge soundness} 
    We argue knowledge soundness of the poly-IOP in AGM. The proof sketch is as follows. Given that the verifier $\mathcal{V}$ accepts the proof, the prover $\mathcal{P}$ must send the commitments that can pass the pairing checks in step 6 of the protocol, which implies that $\mathcal{P}$ must also send the corresponding coefficients of the polynomials $h_1(x), h_2(x),Q(x),f_i(x)$ for $i\in[1,N]$ such that
    \[h_1(x)(x-\zeta)=Q(x)-Q(\zeta)+\sum_1^N\gamma^k(f_k(x)-f_k(\zeta))\] and 
    \[
    h_2(x)(x-\omega\zeta)=\sum_1^{N-1}\gamma^k(f_k(x)-f_k(\omega\zeta))
    \]
    The first equation implies the openings at $\zeta$ must be true for $Q(x)$ and $f_i(x)$ for $i\in[1,N]$. Otherwise, we can construct a polynomial $R(x)$, which is
    \[
    Q(x)-Q(\zeta)+\sum_1^N\gamma^k(f_k(x)-f_k(\zeta)) - h_1(x)(x-\zeta)
    \] 
    And we can break the DLOG assumption since it is easy to find the root of $R(x)$, which is the trapdoor in the KZG commitment. We skip the argument for the second equation as it is similar. Finally, since the extractor now has all $f_i(x)$ for $i\in[1,N]$ in AGM, the extractor can recover the witness, which are the evaluations $f_i(\omega^j),\; \forall j\in[0,n-1], \forall i\in[1,N]$.

    \item \textbf{Zero knowledge}
    The simulator is defined as follows:\\
    $\mathcal S(x,\texttt{srs},\pi)$:
    \begin{enumerate}[leftmargin=*]
    \item Sample $f_{1,x},\ldots,f_{n,x},f_{1,\omega x},\ldots,f_{n,\omega x}\leftarrow\mathbb F$.
    \item Calculate $Q_x= \frac{1}{x^n-1} R(f_{1,x},\ldots,f_{1,x},f_{1,\omega x},\ldots,f_{1,\omega x})$.
    \item Send $[f_{1,x}]_1,\ldots,[f_{n,x}]_1,[Q_x]_1$ to $V^*$.
    \item Receive $\zeta$ from $V^*$.
    \item Sample $f_{1,\zeta},\ldots,f_{n,\zeta},f_{1,\omega\zeta},\ldots,f_{n,\omega\zeta}\leftarrow\mathbb F$.
    \item Calculate $Q_\zeta= \frac{1}{\zeta^n-1} R(f_{1,\zeta},\ldots,f_{1,\zeta},f_{1,\omega\zeta},\ldots,f_{1,\omega\zeta})$.
    \item Send $f_{1,\zeta},\ldots,f_{n,\zeta},f_{1,\omega\zeta},\ldots,f_{n,\omega\zeta},Q_\zeta$ to $V^*$.
    \item Receive $\gamma$ from $V^*$.
    \item Calculate $h_{1,x} = \frac{1}{x-\zeta} \left(Q_x - Q_\zeta + \sum_{k=1}^N \gamma^k [f_{k,x} - f_{k,\zeta}]\right)$ and $h_{2,x} = \frac{1}{x-\omega\zeta} \sum_{k=0}^{N-1} \gamma^{k-1} [f_{k,x} - f_{k,\omega\zeta}]$.
    \item Send $[h_{1,x}]_1$ and $[h_{2,x}]_1$ to $V^*$.
    \end{enumerate}
    \begin{proof}
    The transcript between $\mathcal S(x,\texttt{srs},\pi)$ and $\mathcal{V}^*(\texttt{srs},\pi)$ is indistinguishable from the transcript between
    $\mathcal P(\texttt{srs},\pi,w)$ and $\mathcal{V}^*(\texttt{srs},\pi)$.
    According to Lemma 1 in \cite{sefranek2024simulate}, $f_i(x),f_i(\zeta),\allowbreak f_i(\omega\zeta)$ are indistinguishable from random, which is what is generated by $\mathcal S$.
    Also, the values of $[Q(\tau)]_1,Q(\zeta),[h_1(\tau)]_1, \allowbreak [h_2(\tau)]_1$ returned by $\mathcal P$ are indistinguishable from the values returned by $\mathcal S$ since these values are deterministic given the earlier values in the transcript.
    \end{proof}
\end{enumerate}
\renewcommand{\thesection}{C}
\section{Proof for knowledge soundness of the generalized Mira}

The accumulation scheme based on algebraic checks satisfies knowledge soundness due to the applicability of the Fiat-Shamir transform to an underlying $(d+1)$-special sound protocol \cite{bunz2023protostar}. This transformation enables us to argue that the accumulation scheme inherits knowledge soundness.

We analyze the public-coin interactive protocol 
$\Pi_{\mathsf{I}} = (\mathcal{P}_{\mathsf{I}}(\text{acc}, \text{acc}'), \allowbreak \mathcal{V}_{\mathsf{I}}(\text{acc}.\mathsf{x}, \text{acc}'.\mathsf{x})),$ where $\mathcal{P}_{\mathsf{I}}$ sends $\text{pf} = [E_j]_{j=1}^{d-1} \in \mathbb{G}^{d-1}$ (as computed by $\mathcal{P}_{\mathsf{acc}}$) to $\mathcal{V}_{\mathsf{I}}$. The verifier then provides a random challenge $\alpha \in \mathbb{F}$, and the prover $\mathcal{P}_{\mathsf{I}}$ responds with $\text{acc}''$ calculated by $\mathcal{P}_{\mathsf{acc}}$. The verifier $\mathcal{V}_{\mathsf{I}}$ accepts if $\mathcal{D}_{\mathsf{acc}}(\text{acc}'')$ and $\mathcal{V}_{\mathsf{acc}}(\text{acc}.\mathsf{x}, \text{acc}'.\mathsf{x}, \text{pf}, \text{acc}''.\mathsf{x})$ pass using $\alpha$.

We now demonstrate that $\Pi_{\mathsf{I}}$ is $(d+1)$-special-sound. Let $T$ denote the $d+1$ accepting transcripts for $\Pi_{\mathsf{I}}$ where
$\{T_i := (\text{acc}.\mathsf{x}, \text{acc}'.\mathsf{x}; \allowbreak \text{acc}''_i, \text{pf}_i)\}_{i=1}^{d+1}$.
From these transcripts, we construct an extractor $\mathsf{Ext}_{\text{acc}}$ that can extract a valid witness.

For each $i \in [d+1]$, $\text{pf}_i = \text{pf} = [E_j]_{j=1}^{d-1}$ and \\
$\text{acc}''_i = (\mu''_i, \mathsf{pi}'', [C''_{i,j}]_{j=1}^k, [r_{i,j}]_{j=1}^{k-1}, E''_i, [m''_{i,j}]_{j=1}^k)$.

Since all transcripts are accepting, both $\mathsf{V}_{\text{acc}}$ and $\mathsf{D}_{\text{acc}}$ succeed. Therefore, we have for all $i \in [d+1]$
\[
\mathsf{Commit}(\mathsf{ck}, e''_i) = E''_i = \text{acc}'.E + \sum_{j=1}^{d-1} \alpha_i^j E_j +\alpha_i^d \cdot \text{acc}.E \quad
\]
where
\[
e''_i := \sum_{j=0}^d \mu''^{d-j}_i f^{\mathsf{Vsps}}_j(\mathsf{pi}'', [m''_{i,j}]_{j=1}^k, [r_{i,j}]_{j=1}^{k-1}).
\]

Following the approach of \cite{bunz2023protostar}, we use the Vandermonde matrix formed by the challenges $\alpha_1, \ldots, \alpha_{d+1}$ to derive $e, e', [e_j]_{j=1}^{d-1}$ such that $E_j = \mathsf{Commit}(\mathsf{ck}, e_j)$, $\text{acc}.E = \mathsf{Commit}(\mathsf{ck}, e)$, and $\text{acc}'.E = \mathsf{Commit}(\mathsf{ck}, e')$ from the relations above. Thus,
\[
e''_i = e' + \sum_{j=1}^{d-1} \alpha_i^j e_j + \alpha_i^d\cdot e \quad \text{for all } i \in [d+1].
\]

Furthermore, using just two challenge values $(\alpha_1, \alpha_2)$, $\mathsf{Ext}_{\text{acc}}$ can compute
$\text{acc}.m_j = (\text{acc}''_1.m_j - \text{acc}''_2.m_j) / (\alpha_1 - \alpha_2) \quad \forall j \in [k]$.

Then,
$\text{acc}'.m_j = \text{acc}''_1.m_j - \alpha_1 \cdot \text{acc}.m_j \, \forall j \in [k]$,
ensuring $\text{acc}.C_j = \mathsf{Commit}(\mathsf{ck}, \text{acc}.m_j)$ and $\text{acc}'.C_j = \mathsf{Commit}(\mathsf{ck}, \allowdisplaybreaks \text{acc}'.m_j)$.

If, for some other challenge and any $j$, it holds that $\text{acc}''.m_j \ne \alpha \cdot \text{acc}.m_j + \text{acc}'.m_j$, this would violate the binding property of the commitment scheme $\mathsf{cm}$—an event that occurs only with negligible probability under standard cryptographic assumptions.

Otherwise, this implies the following degree-$d$ polynomial:
% \todo{}{can we define variables to make the second line shorter}
\begin{align*} 
&p(X) = \sum_{j=0}^d (X \cdot \text{acc}.\mu + \text{acc}'.\mu)^{d-j} \cdot \\ &f_j^{\mathsf{Vsps}}(X \cdot \mathsf{pi} + \mathsf{pi}', [X \cdot =m_i + m_i']_{i=1}^k,\; [X \cdot r_i + r_i']_{i=1}^{k-1}) \\& - e' - \sum_{j=1}^{d-1} e_j X^j - e\cdot X^d
\end{align*}
is zero at $d+1$ distinct points $(\alpha_1, \ldots, \alpha_{d+1})$, implying it is identically zero. The constant term of this polynomial 
% \[
% \sum_{j=0}^d \text{acc}'.\mu^{d-j} \cdot f_j^{\mathsf{Vsps}}(\text{acc}'.\mathsf{pi},\; [\text{acc}'.m_i]_{i=1}^k,\; [\text{acc}'.r_i]_{i=1}^{k-1}) - e',
% \]
% which 
being zero implies $\mathcal{D}(\text{acc}') = 0$ and the degree-$d$ coefficient
% \[
% \sum_{j=0}^d \text{acc}.\mu^{d-j} \cdot f_j^{\mathsf{Vsps}}(\text{acc}'.\mathsf{pi},\; [\text{acc}.m_i]_{i=1}^k,\; [\text{acc}.r_i]_{i=1}^{k-1}) - e,
% \]
being zero implies $\mathcal{D}(\text{acc}) = 0$. Hence, $\mathsf{Ext}$ successfully extracts a valid witness $(\text{acc}.\mathsf{w}, \text{acc}'.\mathsf{w})$, establishing that $\Pi_{\mathsf{I}}$ is $(d+1)$-special-sound. By Lemma 1 from \cite{bunz2023protostar}, the Fiat-Shamir transform $\Pi_{\mathsf{AS}} = \mathsf{FS}[\Pi_{\mathsf{I}}]$ results in a NARK for $\mathcal{R}_{\text{acc}}$ with knowledge soundness bounded by
$(Q+1)(d+1)/|\mathbb{F}| + \text{negl}(\lambda)$
where $Q$ is the number of oracle queries performed by the attacker.

Finally, using techniques similar to those in \cite{beal2024mira}, we can show that protocols relying on algebraic tests (e.g., pairings) also achieve knowledge soundness by using extractors in \cite{beal2024mira} to transform them into algebraic check-based protocols as done here.
\end{document}